\documentclass[iop,revtex4]{emulateapj}
\usepackage{lineno}

\usepackage{lscape}

\newcommand{\msun}{${\cal M}_\odot$}
\newcommand{\kms}{km~s$^{-1}$}
\newcommand{\masyr}{mas~yr$^{-1}$}

\begin{document}

\renewcommand{\topfraction}{1.0}
\renewcommand{\bottomfraction}{1.0}
\renewcommand{\textfraction}{0.0}

\shorttitle{Speckle observations and orbits of multiple stars}
\shortauthors{Tokovinin et al.}

\title{Speckle observations and orbits of multiple stars}

\author{Andrei Tokovinin}
\affil{Cerro Tololo Inter-American Observatory, Casilla 603, La Serena, Chile}
\email{atokovinin@ctio.noao.edu}

\author{Mark E. Everett}
\affil{National Optical Astronomy Observatory, 950 North Cherry Avenue, Tucson, AZ 85719, USA}
\email{everett@noao.edu}

\author{Elliott P. Horch\footnote{Adjunct Astronomer, Lowell Observatory} }
\affil{Department of Physics, Southern Connecticut State University, 501 Crescent Street, New Haven, CT 06515, USA}
\email{horche2@southernct.edu}

\author{Guillermo Torres}
\affil{Center for Astrophysics | Harvard \& Smithsonian, 60 Garden Street, Cambridge, MA 02138, USA}
\email{gtorres@cfa.harvard.edu}

\author{David W. Latham}
\affil{ Center for Astrophysics | Harvard \& Smithsonian, 60 Garden Street, Cambridge, MA 02138, USA}
\email{dlatham@cfa.harvard.edu},

\begin{abstract}
We  report  results of  speckle-interferometric  monitoring of  visual
hierarchical systems using the  newly commissioned instrument NESSI at
the 3.5-m  WIYN telescope.  During  one year, 390 measurements  of 129
resolved  subsystems were  made, while  some targets  were unresolved.
Using our astrometry and archival  data, we computed  36 orbits (27 for
the first  time).  Spectro-interferometric  orbits of seven  pairs are
determined by combining positional measurements with radial velocities
measured,   mostly,   with  the   Center   for  Astrophysics   digital
speedometers. For  the hierarchical systems HIP~65026  (periods 49 and
1.23 years)  and HIP 85209 (periods  34 and 1.23  years) we determined
both  the inner  and  the  outer orbits  using  astrometry and  radial
velocities   and   measured   the   mutual   orbit   inclinations   of
11\fdg3$\pm$1\fdg0 and  12\fdg0$\pm$3\fdg0, respectively.  Four bright
stars  are  resolved for  the  first time;  two  of  those are  triple
systems.  Several  visual subsystems  announced in the  literature are
shown to be spurious.  We  note that subsystems in compact hierarchies
with outer  separations less than 100  au tend to  have less eccentric
orbits compared to wider hierarchies.
\end{abstract} 

\section{Introduction}
\label{sec:intro}

We know that  the majority of  multi-planet  systems are co-planar
  \citep{Fabrycky2014},  as well  the planets  found  around eclipsing
  binaries  \citep[e.g. Kepler-1647,][]{Kostov2016}.   Low-mass triple
systems with  outer separations  of the order  of $\sim$50 au  or less
also have a  strong alignment trend, even though  the typical relative
angles between orbital  planes are still substantial, $\sim$30$^\circ$
\citep{Tok2017}.  Formation  of companions in  a massive circumstellar
disc, their subsequent  growth and inward migration seem  to match the
prevalent      architecture     of      compact     ``planetary-like''
(i.e. quasi-coplanar) triple systems.   One of the natural outcomes of
this scenario  are close (eclipsing) binaries surrounded  by a remnant
disk, from which the circumbinary planets condense, as demonstrated by
the  {\it  Kepler} discovery  of  such  systems.   Recent modeling  by
\citet{Moe2018} shows that  migration in a disc is  also necessary for
explaining the  observed fraction of solar-type  close binaries, while
the  alternative  channel  of  close-binary  formation  via  dynamical
evolution of misaligned triple systems, also relevant for formation of
hot Jupiters \citep{FT2007}, is not efficient enough.

The distribution  of  relative  inclinations   in  triple  systems  is  an
essential input  to those studies. Orbits  of binary and  triple systems
contain  a fossil  record of  the conditions  prevailing  during their
formation.  While migration in a disk produces quasi-coplanar systems, 
misaligned systems diagnose alternative formation mechanisms. For example,
  frequent dynamical  interactions between  stars  in dense
clusters or accretion of  misaligned gas produce multiple systems with
misaligned orbits, binaries with  misaligned stellar spins and, later,
planetary  systems that  are  not confined  to  one plane.

Although stellar triples have been  known for a long time, advances in
the  observational techniques have only  recently given  access to  the most
interesting close and fast systems with spatial scales of $\sim$10 au,
matching the typical size of discs and planetary systems.  For some of
those, it has  been possible to determine both  inner and outer orbits
and  their  relative  orientation \citep{planetary,TL2017}.   However,
much  work  remains  to  be done.   Ongoing  high-resolution  surveys,
particularly in  the solar neighborhood, continue  to discover compact
hierarchical   systems   \citep{Horch2019}.    Follow-up   astrometric
monitoring on a time scale from  years to decades is needed to map the
orbital motion and to define the architecture of those systems.  These
data, in turn, will provide insights on the formation history of stars
and planets.

The  highly efficient  speckle program  at the  Southern Astrophysical
Research Telescope (SOAR) assigns top priority to the study of orbital
motions   in  resolved   triple  systems   \citep[see][and  references
  therein]{SOAR2018}.  As  a result, tens of triples  already have new
or  updated   accurate  orbits  while  monitoring   of  other  systems
continues.  We are extending this effort to the northern sky using the
recently  commissioned NESSI  instrument on  the 3.5-m  WIYN telescope
\citep{NESSI}.  

Our   targets  were  selected   from  the   current  version   of  the
Multiple-Star  Catalog \citep{MSC}.   We observed  resolved northern subsystems
with   unknown  orbits  and estimated  orbital
periods shorter  than 100  yr.  This speckle  program at  WIYN pursues
three main goals.
\begin{itemize}
\item
Determination  or   improvement  of  visual  orbits   for  members  of
hierarchical  systems, especially  those with  short  periods.  Orbits
contribute to  the statistics  of periods, relative  inclinations, and
eccentricities.

\item
Positional measurements and relative photometry of subsystems with 
periods of a few decades for future orbit determinations. 

\item
Verification  of  uncertain  subsystems.  Identification  of  spurious
triple  stars  cleans  the  statistics  and saves  efforts  for  their
future monitoring.

\end{itemize}

In this  paper we  report speckle observations  at the  WIYN telescope
conducted during  one year, from  2018 February to 2019  January.  The
instrument,  observing procedure,  and data  reduction are  covered in
Section~\ref{sec:obs}.  The results  (new measurements and orbits) are
presented  in Section~\ref{sec:res}.  Section~\ref{sec:sum} summarizes
our work.


\section{Observations}
\label{sec:obs}

\subsection{The instrument}

The  speckle camera at  the WIYN  telescope is  called NESSI  for {\it
  NASA-NSF  Exoplanet  Observational  Research  (NN-EXPLORE)  Exoplanet
  Stellar Speckle Imager.}  This  is a dual-channel imaging instrument
\citep{NESSI}.  The  beam arriving  from the telescope  is collimated,
divided  into  two color  channels  by  a  dichroic beamsplitter,  and
re-focused on two identical electron-multiplication CCD cameras with a
magnification of 1.5$\times$ or a pixel scale of 18.2 mas. Wavelengths
longer  than 673\,nm  pass through  the dichroic  to the  red channel,
while  shorter wavelengths are  reflected to  the blue  channel.  Both
channels  have additional  filters.  Here  we used  the  562.3/43.6 nm
filter in  the blue channel and  the 716.0/51.5 nm in  the red channel
(some  observations  were  obtained   instead  in  the  832.0/40.4  nm
filter). NESSI does not have an atmospheric dispersion corrector.  The
position  angle  on the  sky  is  maintained  by the  WIYN  instrument
rotator.  In  the blue  channel,  the  image  is oriented  with  North
pointing to the  left and East pointing down. In  the red channel, the
horizontal  orientation is  inverse (North  pointing  right).  Further
information on NESSI is provided by \citet{NESSI}.

\subsection{Observing procedure}

On  each target,  we  registered sub-frames  of 256$\times$256  pixels
(4\farcs8 on the  sky) with a short exposure  of 40\,ms.  The electron
multiplication gain was adjusted  depending on the star brightness.  A
sequence of 1000  such frames (the data cube) is  recorded in the FITS
file.  Auxiliary  information (target name,  coordinates, filters, and
time  of observation)  is kept  in the  text files    of observing
  logs, one per night.  We recorded from 2 to 8 sequential data cubes
of the object  and one data cube of the  reference star, selected from
the Bright Star Catalog in the vicinity of the target. The bright star
is used as a reference in the data processing.

Forty hours of service observations  were allocated to this program by
the  NASA-WIYN TAC  (programs  18A-0122 and  18B-0008).  Our  objects,
reference stars, and calibration  binaries were observed together with
other programs. The data used  in this paper have been obtained during
33 nights  in the period from 2018  February 2 to 2019  January 30.  A
total of 2057  data cubes were recorded, amounting to  23 hours of the
open-shutter time.

\subsection{Data processing and calibration}
\label{sec:dat}

All NESSI  data cubes are  normally processed using the  speckle image
reconstruction pipeline \citep{NESSI}. However, for multiple stars the
image  reconstruction is  not  strictly necessary  (images from the NESSI pipeline are used here for illustration). 
We adaptated the  SOAR  speckle pipeline  \citep[see][]{TMH10,HRCAM}
written in IDL.   The most computer-intensive part of  the pipeline is
the calculation  of the power  spectra.  After reading the  image cube
into  the memory,  the code  first  determines the  centroids in  each
frame, thresholded at 10 counts above the background (the threshold is
less than one electron and  therefore does not distort the linearity).
Frames where the star is too  close to the edge are rejected.  Then the
power spectrum is  accumulated, as well as re-centered  images and the
shift-and-add  (SAA) images centered  on the  brightest pixel  in each
frame.  The  orientation of  images in the  red channel is  changed to
match the blue channel.

Further data processing uses only the two-dimensional images and power
spectra.  It  is organized by means  of the data  structure that keeps
all relevant parameters and the  processing results for each data cube
and each  channel, i.e. from 4 to  16 records for a  given target and,
typically,  two  records  for   the  reference  star.   The  essential
information is ingested from the  text files of the observing logs and
complemented  by  calculation of  additional  parameters  such as  the
zenith distance  and parallactic angle.  Results of  the data analysis
(e.g.   the signal  level  in  the power  spectrum,  estimates of  the
resolution and detection limits,  and binary-star parameters) are also
kept  in this  structure.   Final results  are  obtained by  averaging
similar  data on  a  given target,  separately  for the  blue and  red
channels.  A more detailed description of the SOAR speckle pipeline can
be found in \citet{HRCAM}.

\begin{figure}
\epsscale{1.1}
\plotone{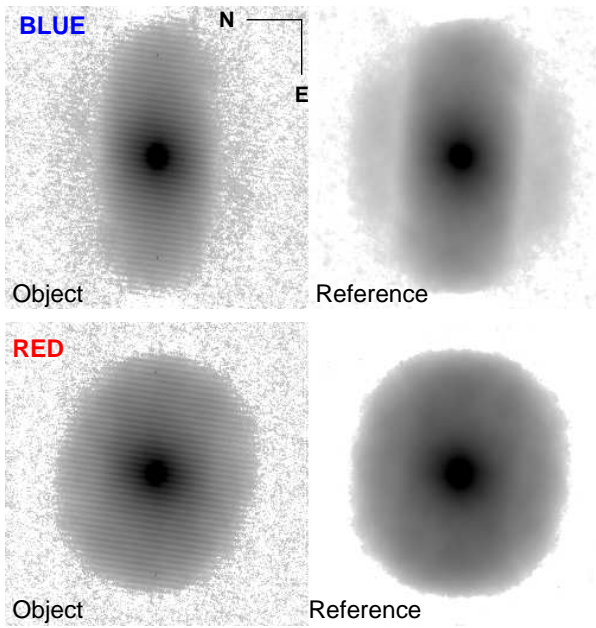}
\caption{Example  of the  power  spectra  in the  blue  (top) and  red
  (bottom) channels of NESSI. The  object (on the left) is a 0\farcs91
  binary BU~403  (WDS J04257$-$0214),  evidenced by the  finely spaced
  horizontal fringes. Power spectra of the reference star are shown on
  the  right. The  zero spatial  frequency is  at the  center  of each
  frame, the power spectra are rendered in negative logarithmic scale.
  The atmospheric dispersion of 4.6 and 1.3 pixels in the blue and
  red channels, respectively (zenith distance 34\fdg2)  reduces the
  resolution in the North-South direction.
\label{fig:power-sp} }
\end{figure}

Parameters  of   binary  stars  (separation   $\rho$,  position  angle
$\theta$,  and  magnitude difference  $\Delta  m$)  are determined  by
modeling the  observed power  spectrum as a  product of  the reference
spectrum  and  the  binary-star  fringes  \citep{TMH10}.   Uncorrected
atmospheric  dispersion  damps  the  power spectrum  in  the  vertical
direction,  as illustrated  in  Figure~\ref{fig:power-sp}.  The  NESSI
filters have  a rectangular bandpass,  hence the damping factor  has a
shape of  the ${\rm  sinc}^2(x)$ function with  side-lobes.  The  loss of
resolution is  particularly important in the  blue channel, offsetting
the greater resolving power at   shorter wavelengths. As the object
and  reference are  observed close  to  each other,  the   model
adequately  accounts  for  the  dispersion. Triple  stars  are  fitted
similarly.

We compared  the relative  scale and orientation  in the blue  and red
channels using  24 binaries with separations larger  than 0\farcs5 and
$\Delta m < 3$ mag. The red camera has been removed between some runs,
and, as a result, the  total range of relative angular offsets between
the channels is 1\fdg2. On the other hand, the difference in the scale
factor is  quite consistent:  pixels in the  red camera are  larger by
0.8\%.  We corrected  the astrometry in the red  channel for scale and
angle offsets to match the blue channel.

Twelve relatively wide binaries observed  at WIYN in 2018 February and
June  belong  to the  list  of  SOAR  calibrators with  accurately
modeled  motion  \citep{HRCAM}.  The   mean  angular  offset  of  WIYN
measurements  of those stars  relative to  their models  is $-$0\fdg15
with an rms  scatter of 0\fdg17; their separations  match the model to
within 0.3\%. We corrected for the angular offset and left the separations
as measured, thus confirming the nominal pixel scale of the NESSI blue
channel, 18.2\,mas. We  assume here that the calibration  of angle and
pixel scale of NESSI remained stable throughout the whole year. A more
detailed study of the  calibration would require repeated measurements
of a  larger number of wide  calibration pairs, not  available in this
data set. We  note that residuals of our  measurements to good-quality
orbits are  small, confirming independently the  absence of systematic
errors at the level of $<$0\fdg5 in angle and $<$0.5\% in scale.

\subsection{Radial velocities}

Radial  velocities  for several  systems  have  been  obtained at  the
Harvard-Smithsonian  Center  for   Astrophysics  (CfA)  using  several
instruments  on   different  telescopes.   The   Digital  Speedometers
\citep[see][]{Latham:1992} were  used on 1.5  m telescopes at  the Oak
Ridge Observatory  (Harvard, Massachusetts)  and the Fred  L.\ Whipple
Observatory  (Mount  Hopkins, Arizona),  and  on  the Multiple  Mirror
Telescope with  equivalent aperture of  4.5 m (also on  Mount Hopkins)
before its conversion  to a monolithic 6.5 m  telescope.  In these
  instruments,   intensified    photon-counting   Reticon   detectors
delivered   a  single   echelle   order  45~\AA\   wide  centered   at
5187~\AA\ (featuring  the \ion{Mg}{1}~b triplet) at  a resolving power
of 35,000. 

We   also  used   the  Tillinghast   Reflector   Echelle  Spectrograph
\citep[TRES;][]{Szentgyorgyi:2007, Furesz:2008} attached  to the 1.5 m
telescope  on  Mount  Hopkins,   which  covers  the  wavelength  range
3900--9100~\AA\ in 51 orders at a resolving power of 44,000. Most
  data  come from  the  CfA spectrometers,  while  TRES contributed  34
  observations, mostly of HIP 85209.

Radial  velocities were measured  by cross-correlation  using suitable
synthetic or observed templates centered on the \ion{Mg}{1}~b triplet.
For systems showing double or triple lines we used the two-dimensional
or three-dimensional correlation techniques TODCOR \citep{Zucker:1994}
and  TRICOR \citep{Zucker:1995},  with  a separate  template for  each
stellar  component.  The  RVs  are given  below in  Table~\ref{tab:rv}
together with their residuals from  the orbits.  The CfA RVs are given
on the native  instrument system (a correction of  +0.14 \kms would be
needed to bring  them to the IAU system).  The  RVs measured with TRES
have been corrected to the same  zero point as the velocities from the
CfA Digital Speedometers,  so they also need to  be corrected by +0.14
\kms to put them on the IAU system.

\section{Results}
\label{sec:res}

\begin{figure}
\epsscale{1.1}
\plotone{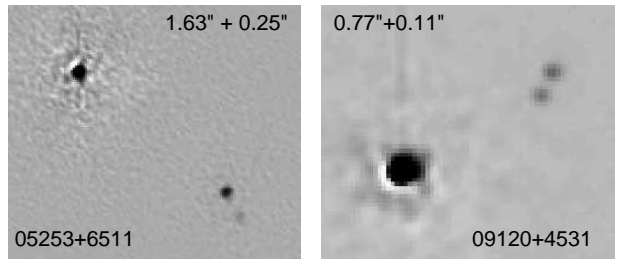}
\caption{Reconstructed images of two triple systems in the red channel.
Left: WDS J05253+6511 (HDS 711 A,Ba and BAG~17 Ba,Bb) observed in 2018.1.
Right: WDS J09120+4531 (YSC~91 A,Ba and Ba,Bb) observed in
2019.1. Separations of the wide and close pairs are indicated. 
\label{fig:images} }
\end{figure}

\subsection{Speckle measurements}

The results  (measurements of resolved pairs  and non-resolutions) are
presented  in  almost the  same  format  as  in the  papers  reporting
observations at  SOAR.  The long tables  are published electronically;
here   we  describe   their  content.   Figure~\ref{fig:images}  shows
.images  of two triple systems   reconstructed by the
  NESSI  pipeline. Although  the measurements  are obtained  from the
power  spectra, the  reconstructed images  of selected  triple systems
illustrate the imaging capability of NESSI. 

\begin{deluxetable}{ l l  l l }
\tabletypesize{\scriptsize}
\tablewidth{0pt}
\tablecaption{Measurements of multiple stars at WIYN
\label{tab:measures}}
\tablehead{
\colhead{Col.} &
\colhead{Label} &
\colhead{Format} &
\colhead{Description, units} 
}
\startdata
1 & WDS    & A10 & WDS code (J2000)  \\
2 & Disc.  & A16 & Discoverer code  \\
3 & Other  & A12 & Alternative name \\
4 & RA     & F8.4 & R.A. J2000 (deg) \\
5 & Dec    & F8.4 & Declination J2000 (deg) \\
6 & Epoch  & F9.4 & Julian year  (yr) \\
7 & Filt.  & A3 & Filter ($\lambda$ in nm) \\
8 & $N$    & I2 & Number of averaged cubes \\
9 & $\theta$ & F8.1 & Position angle (deg) \\
10 & $\rho \sigma_\theta$ & F5.1 & Tangential error (mas) \\
11 & $\rho$ & F8.4 & Separation (arcsec) \\
12 &  $\sigma_\rho$ & F5.1 & Radial error (mas) \\
13 &  $\Delta m$ & F7.1 & Magnitude difference (mag) \\
14 & Flag & A1 & Flag of magnitude difference\tablenotemark{a} \\
15 & (O$-$C)$_\theta$ & F8.1 & Residual in angle (deg) \\
16 & (O$-$C)$_\rho$ & F8.3 & Residual in separation (arcsec) \\
17  & Ref. & A8   & Orbit reference\tablenotemark{b} 
\enddata
\tablenotetext{a}{Flags: 
q -- the quadrant is determined; 
* -- $\Delta m$ and quadrant from average image; 
: -- noisy data. }
\tablenotetext{b}{References to VB6 are provided at
  \url{http://ad.usno.navy.mil/wds/orb6/wdsref.txt} }
\end{deluxetable}

Table~\ref{tab:measures} lists 390  measurements of 129 resolved pairs
and subsystems,  including new discoveries.  The  pairs are identified
by   their  Washington   Double  Star   (WDS)  codes   and  discoverer
designations adopted  in the  WDS catalog \citep{WDS},  as well  as by
alternative  names in  column  (3), mostly  from  the {\it  Hipparcos}
catalog.  Equatorial  coordinates for the  equinox and epoch  J2000 in
degrees are given  in columns (4) and (5)  to facilitate matching with
other  catalogs and  databases.   Column (6)  lists the  dates of
  observation in Julian years, column  (7) the filters, column (8) the
  numbers of  averaged measurements  from individual data  cubes.  The
  measurements  (position angles  $\theta$, errors  in  the tangential
  direction  $\rho \sigma_\theta$,  separations $\rho$,  radial errors
  $\sigma_\rho$, and  magnitude differences  $\Delta m$) are  given in
  columns  (9)  to (13),  respectively.    In  the case  of  multiple
systems,  the position  measurements  and their  errors and  magnitude
differences refer  to the individual pairings  between components, not
to their photo-centers.  

We list the internal errors  derived from the power spectrum model and
from  the difference between  the measurements  in several  data cubes
\citep{TMH10} and do not  include systematic errors.  The median error
is 0.4\,mas, and 94\% of errors are less than 4\,mas.  The real errors
are somewhat  larger, especially for difficult  pairs with substantial
$\Delta  m$  and/or  with   small  separations  (an  example  of  such
challenging    pair    12576+3514   Aa,Ab    is    given   below    in
Section~\ref{sec:sb}).   We checked the  estimated internal  errors by
comparing the measurements in the  blue and red channels and computing
$\chi^2_\rho  =   (\rho_{b}  -  \rho_{r})^2   /(\sigma^2_{\rho,  b}  +
\sigma^2_{\rho, r})$ and an  analogous $\chi^2_\theta$ using 114 pairs
measured in  both channels simultaneously. The  median $\chi^2$ values
are 2.1  and 1.6  in $\rho$ and  $\theta$, respectively, with  a large
scatter; $\chi^2$  do not correlate  with $\rho$ and $\Delta  m$.  The
scatter  is reduced  and the  $\chi^2$ medians  approach  the expected
value of  one if the  instrumental errors of  0.3 and 0.15 mas  in the
radial  and  tangential  directions  are added  quadratically.   These
numbers  quantify  the typical  difference  between  the internal  and
external errors of our astrometry.

The  flags in column  (14) indicate  cases when  the true  quadrant is
determined  from  the SAA  images  (otherwise  the  position angle  is
measured  modulo  180\degr), when  the  photometry  of  wide pairs  is
derived from the long-exposure images (this reduces the bias caused by
speckle anisoplanatism) and when the data are noisy or the resolutions
are tentative.  For  binary stars with known orbits,  the residuals to
the latest orbit and its reference are provided in columns
(15)--(17). The orbits computed or updated here are denoted as WIYN2019.

Non-resolutions (including  those of reference stars)  are reported in
Table~\ref{tab:single}.  Its first columns  (1) to  (8) have  the same
meaning and  format as  in Table~\ref{tab:measures}. Column  (9) gives
the minimum resolvable  separation when pairs with $\Delta  m < 1$ mag
are detectable. It  is computed from the maximum  spatial frequency of
the useful signal  in the power spectrum and is  normally close to the
formal diffraction  limit $\lambda/D$. The following  columns (10) and
(11) provide the indicative  dynamic range, i.e. the maximum magnitude
difference at separations of 0\farcs15 and 1\arcsec, respectively. 

\begin{deluxetable}{ l l  l l }
\tabletypesize{\scriptsize}
\tablewidth{0pt}
\tablecaption{Unresolved stars 
\label{tab:single}}
\tablehead{
\colhead{Col.} &
\colhead{Label} &
\colhead{Format} &
\colhead{Description, units} 
}
\startdata
1 & WDS    & A10 & WDS code (J2000)  \\
2 & Disc.  & A16 & Discoverer code  \\
3 & Other  & A12 & Alternative name \\
4 & RA     & F8.4 & R.A. J2000 (deg) \\
5 & Dec    & F8.4 & Declination J2000 (deg) \\
6 & Epoch  & F9.4 & Julian year  (yr) \\
7 & Filt.  & A3 & Filter \\
8 & $N$    & I2 & Number of averaged cubes \\
9 & $\rho_{\rm min}$ & F7.3 & Angular resolution (arcsec)  \\
10&  $\Delta m$(0.15) & F7.2 & Max. $\Delta m$ at 0\farcs15 (mag) \\
11 &  $\Delta m$(1) & F7.2 & Max. $\Delta m$ at 1\arcsec (mag) 
\enddata
\end{deluxetable}


\subsection{New and updated visual orbits}
\label{sec:orbits}

\begin{figure}
\epsscale{1.1}
\plotone{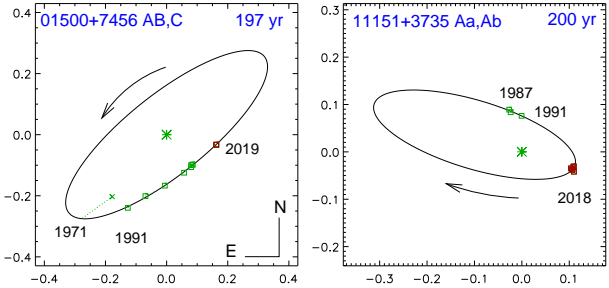}
\caption{Preliminary   (grade    5)   orbits   of    two   long-period
  subsystems. In  these and following plots, the  primary component is
  placed at the coordinate origin. The  axis scale is in arcseconds, North
  is directed up  and East to  the left. Squares denote  the measurements
  (less accurate  measurements are  plotted as crosses);  short dotted
  lines  connect them  to  the corresponding  positions  on the  orbit
  (ellipse). Dates of some measurements are indicated,  the measurements at WIYN are plotted in red.
\label{fig:grade5} }
\end{figure}

Our  measurements  together   with  previous  observations  allow  the
calculation of  orbits for several  subsystems.  We consulted  the WDS
catalog  and the recent literature  for the  existing data.   However, for
many pairs with long periods the available data do not constrain their
orbital elements,  allowing a wide  range of potential  solutions.  We
provide  here preliminary (grade  5) orbits  of such  pairs.  Although
these crude orbits are not suitable for measurements of stellar masses
or  relative orbit  orientation in  triple systems,  their publication
still  makes sense  for improving  the statistics  (even  an uncertain
orbit is better  than no orbit) and planning  future observations.  To
illustrate  this  situation,  two  preliminary  orbits  are  shown  in
Figure~\ref{fig:grade5}. Good  coverage of these  systems with periods
of $\sim$200 years will be  reached only after several more decades of
monitoring.  Meanwhile,  their preliminary orbits  adequately describe
the observed motion and will not need an update  in the near term.

\begin{deluxetable*}{l l cccc ccc cc}    
\tabletypesize{\scriptsize}     
\tablecaption{Visual orbits
\label{tab:vborb}          }
\tablewidth{0pt}                                   
\tablehead{                                                                     
\colhead{WDS} & 
\colhead{Disc.} & 
\colhead{$P$} & 
\colhead{$T$} & 
\colhead{$e$} & 
\colhead{$a$} & 
\colhead{$\Omega_{2000}$ } & 
\colhead{$\omega$ } & 
\colhead{$i$ } & 
\colhead{Grade }  &
\colhead{Reference\tablenotemark{a}} \\
 & & 
\colhead{(yr)} &
\colhead{(yr)} & &
\colhead{(arcsec)} & 
\colhead{(deg)} & 
\colhead{(deg)} & 
\colhead{(deg)} &  & 
}
\startdata
01500$+$7456 & MLR 297 AB,C & 197.5 & 1961.5 & 0.0 & 0.429 & 128.1 & 0.0 & 72.3 & 5 & new \\
02529$+$5300 & A 2906 A,B & 89.6 & 1972.88 & 0.93 & 0.181 & 17.9 & 85.6 & 135.4 & 4 & new \\
        &    & $\pm$4.5 & $\pm$0.79 & fixed & $\pm$0.012 & $\pm$15.8 & $\pm$10.2 & $\pm$4.8&     &  \\
03127$+$7133 & STT 50 A,B & 583 & 2418.5 & 0.428 & 1.866 & 31.0 & 112.2 & 113.9 & 4 & Sca2012b \\
       &    & $\pm$15 & $\pm$15 & $\pm$0.023 & $\pm$0.065 & $\pm$0.9 & $\pm$1.7 & $\pm$0.9&     &  \\
04192$+$6135 & BU 1333 B,C & 180.0 & 2043.6 & 0.67 & 0.238 & 49.3 & 150.0 & 126.9 & 5 & new \\
04480$+$5645 & HDS 617 Aa,Ab & 90 & 2027.8 & 0.87 & 0.503 & 54.5 & 235.9 & 120.0 & 5 & new \\
05237$+$7347 & A 843 A,B & 181.9 & 2036.3 & 0.80 & 0.556 & 27.0 & 230.0 & 86.8 & 4 & new \\
       &    & $\pm$11.6 & $\pm$1.3 & fixed & $\pm$0.023 & $\pm$1.5 & fixed & $\pm$0.4&     &  \\
07211$+$6740 & JNN 55  B,C & 13.97 & 2013.53 & 0.78 & 0.178 & 94.5 & 106.7 & 67.9 & 4 & new \\
       &    & $\pm$1.23 & $\pm$0.90 & fixed & $\pm$0.040 & $\pm$6.5 & $\pm$7.4 & $\pm$7.8&     &  \\
07295$+$3556 & JNN 57 Aa,Ab & 18.4 & 2016.84 & 0.568 & 0.158 & 244.1 & 132.9 & 113.3 & 3 & new \\
       &       & $\pm$2.9 & $\pm$0.66 & $\pm$0.134 & $\pm$0.028 & $\pm$9.6 & $\pm$12.4 & $\pm$8.4&     &  \\
08307$+$4645 & YSC 3 Aa,Ab & 19.32 & 2002.35 & 0.0 & 0.095  & 112.5 & 0.0 & 71.5 & 4 & new \\
       &    & $\pm$0.34 & $\pm$0.18 & fixed & $\pm$0.004  & $\pm$1.7 & fixed & $\pm$1.9&     &  \\
09186$+$2944 & A 221 Ba,Bb & 184 & 1888.1 & 0.90 & 0.591 & 142.3 & 82.7 & 99.2 & 5 & new \\
09354$+$3958 & COU 2084  Aa,Ab & 50.34 & 2012.39 & 0.652 & 0.246 & 84.7 & 137.0 & 112.8 & 3 & new \\
         &    & $\pm$1.91 & $\pm$0.82 & $\pm$0.036 & $\pm$0.012 & $\pm$1.7 & $\pm$6.9 & $\pm$3.4&     &  \\
10454$+$3831 & HO 532  A,C & 326   & 1883   & 0.90 & 3.482 & 154.0 & 95.3 & 96.0 & 5 & Mnt2000a \\
11017$+$3641 & HDS 1574  Aa,Ab & 46.22 & 2013.14 & 0.795 & 0.235 & 169.6 & 337.5 & 116.4 & 4 & new \\
       &    & $\pm$2.57 & $\pm$0.98 & $\pm$0.030 & $\pm$0.011 & $\pm$5.8 & $\pm$14.3 & $\pm$3.2&     &  \\
11017$+$3641 & COU 1422 Ba,Bb & 173   & 1958.0  & 0.35  & 0.56  &  74   & 180   & 39.5  & 5 & new \\
11151$+$3735 & CHR 192 Aa,Ab & 200     & 2019.3   & 0.486 & 0.2224 & 254.2 & 11.2 & 111.7 & 5 & new \\
12089$+$2147 & HDS 1714 Aa,Ab & 235.6   & 2029.2   & 0.80  & 0.5979 & 41.1 & 219.9 & 80.3 & 5 & new \\
14353$+$4302 & LSC 56 Aa,Ab & 11.23 & 2009.85 & 0.30 & 0.0473 & 112.9 & 90.0 & 77.1 & 4 & new \\
        &    & $\pm$1.92 & $\pm$0.88 & $\pm$0.26 & $\pm$0.0057 & $\pm$3.3 & fixed & $\pm$6.0&     &  \\
16147$+$3352 & YSC 152  Ea,Eb & 52.0 & 1994.267 & 0.0 & 0.514 & 214.9 & 0.0 & 60.5 & 5 & Hei1990d \\
19514$+$4044 & COU 2530  A,B & 55.0 & 2022.9 & 0.40 & 0.250 & 159.6 & 38.4 & 74.2 & 4 & new \\
       &    & $\pm$4.8 & $\pm$1.0 & fixed & $\pm$0.007 & $\pm$1.5 & $\pm$8.6 & $\pm$1.3&     &  \\
20599$+$4016 & HDS 2989  Da,Db & 212.7 & 2142.3 & 0.85 & 0.940 & 154.5 & 270.0 & 90.9 & 5 & new \\
21308$+$4827 & A 770  A,B & 146 & 1958.6 & 0.20 & 0.307 & 321.4 & 90.0 & 89.5 & 5 & new \\
22139$+$3943 & MCA 70 Aa,Ac & 125.0 & 2028.8 & 0.82 & 0.324 & 166.1 & 26.0 & 54.6 & 5 & new \\
22359$+$3938 & CHR 112  Aa,Ab & 41.6 & 2022.9 & 0.44 & 0.057 & 119.7 & 107.9 & 69.1 & 3 & new \\
        &    & $\pm$5.4 & $\pm$2.4 & $\pm$0.19 & $\pm$0.007 & $\pm$5.2 & $\pm$5.6 & $\pm$6.4&     & 
\enddata 
\tablenotetext{a}{References: 
Hei1990d -- \citet{Hei1990d}; 
Lin2012b -- \citet{Lin2012b};
Mnt2000a -- \citet{Mnt2000a};
Sca2012b -- \citet{Sca2012b}
}
\end{deluxetable*}

\begin{figure*}
\epsscale{1.1}
\plotone{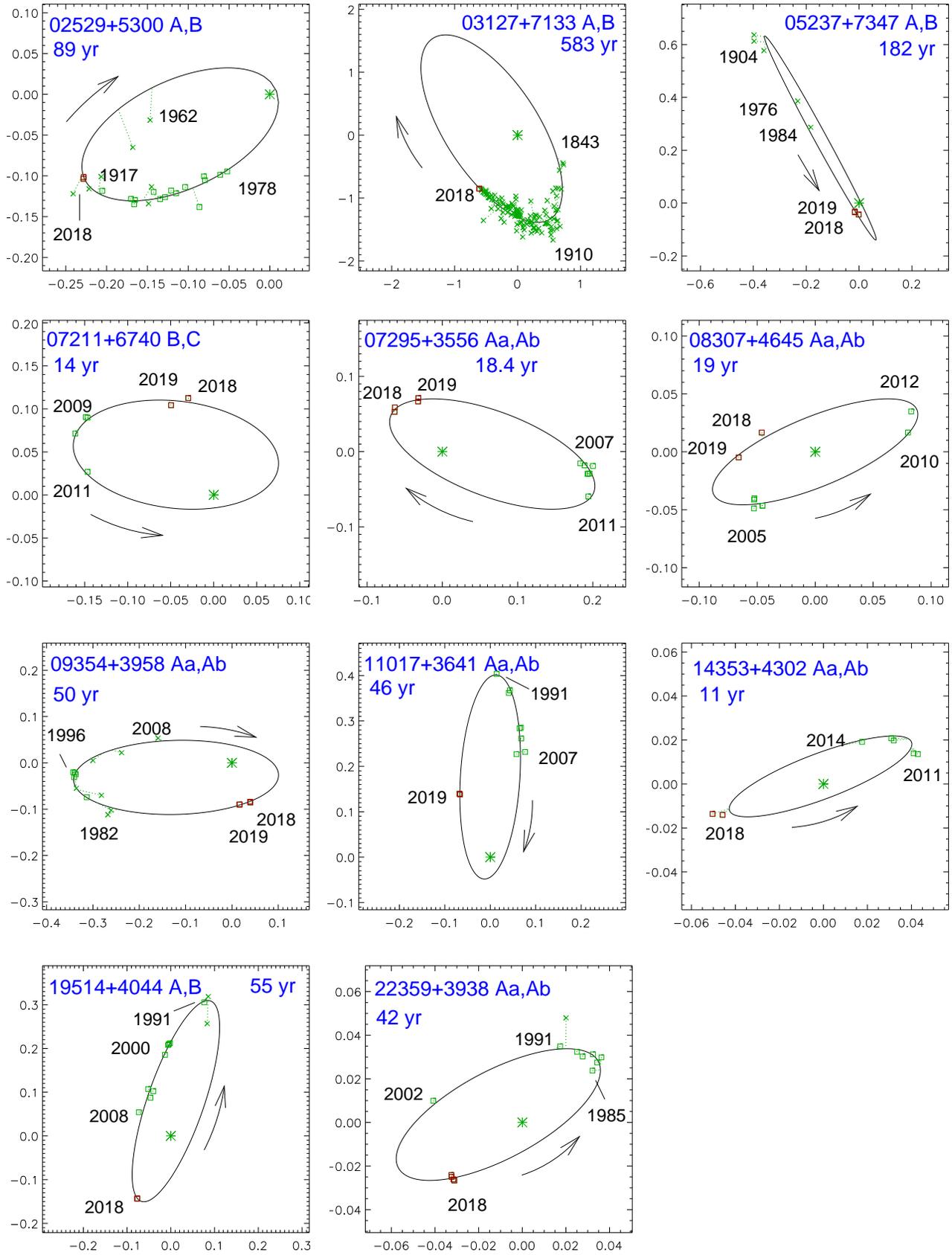}
\caption{Visual orbits    of    grades      3 and    4    (see  the legend    to
  Figure~\ref{fig:grade5}).
\label{fig:grade4} }
\end{figure*}

The  elements   of  visual  orbits  determined  here   are  listed  in
Table~\ref{tab:vborb} in standard  notation.  The position angles were
corrected for  precesssion, so the  nodes $\Omega$ refer to  the J2000
epoch.  The orbits were computed by least-squares fitting with weights
inversely proportional to the adopted measurement errors; the IDL code
{\tt ORBIT} \citep{ORBIT} was used.  The two ultimate columns give the
provisional grade  and the  reference to previously  published orbits,
when available. We do not list formal errors for preliminary orbits of
grade   5.   Orbits   of  grades   4   and  3   are  plotted   in
Figure~\ref{fig:grade4}.

\begin{deluxetable*}{r l l rrr rr l}    
\tabletypesize{\scriptsize}     
\tablecaption{Position measurements and residuals (fragment)
\label{tab:speckle}          }
\tablewidth{0pt}                                   
\tablehead{                                                                     
\colhead{WDS} & 
\colhead{Disc.} & 
\colhead{Date} & 
\colhead{$\theta$} & 
\colhead{$\rho$} & 
\colhead{$\sigma_\rho$} & 
\colhead{(O$-$C)$_\theta$ } & 
\colhead{(O$-$C)$_\rho$ } &
\colhead{Ref.\tablenotemark{a}} \\
 & & 
\colhead{(yr)} &
\colhead{(\degr)} &
\colhead{(\arcsec)} &
\colhead{(\arcsec)} &
\colhead{(\degr)} &
\colhead{(\arcsec)} &
}
\startdata
00541+6626 & YSC 19 Aa,Ab     & 2018.6496 & 134.6 & 0.0800 &  0.0050 &     0.0 &  -0.0024 & W\\
00541+6626 & YSC 19 Aa,Ab     & 2018.6496 & 134.0 & 0.0820 &  0.0050 &    -0.6 &  -0.0004 & W\\
01500+7456 & MLR 297 AB,C      & 1971.5900 & 138.5 & 0.2700 &  0.1500 &     5.0 &  -0.1395 & M\\
01500+7456 & MLR 297 AB,C      & 1991.2500 & 152.0 & 0.2710 &  0.0100 &     1.2 &  -0.0014 & H \\
01500+7456 & MLR 297 AB,C      & 1999.8115 & 161.0 & 0.2129 &  0.0050 &    -6.7 &   0.0208 & s\\
01500+7456 & MLR 297 AB,C      & 2000.8743 & 178.0 & 0.1670 &  0.0050 &     7.2 &  -0.0157 & s \\
01500+7456 & MLR 297 AB,C      & 2007.8215 & 204.7 & 0.1370 &  0.0050 &     4.3 &   0.0006 & s \\
01500+7456 & MLR 297 AB,C      & 2010.7130 & 217.2 & 0.1334 &  0.0050 &    -0.0 &   0.0029 & s 
\enddata 
\tablenotetext{a}{
H: Hipparcos;
h: HST; 
M: visual micrometer measures;
s: speckle interferometry or lucky imaging at other telescopes;
a: adaptive optics;
G: Gaia;
W: speckle interferometry at WIYN.
}
\end{deluxetable*}

\begin{deluxetable*}{r l c rrr l }    
\tabletypesize{\scriptsize}     
\tablecaption{Radial velocities and residuals (fragment)
\label{tab:rv}          }
\tablewidth{0pt}                                   
\tablehead{                                                                     
\colhead{WDS} & 
\colhead{Disc.} & 
\colhead{Date} & 
\colhead{RV} & 
\colhead{$\sigma$} & 
\colhead{(O$-$C)$$ } &
\colhead{Comp.}
 \\
 & & 
\colhead{(HJD $-$2400000)} &
\multicolumn{3}{c}{(km s$^{-1}$)}  &
\colhead{Instr.\tablenotemark{a}}
}
\startdata
00541+6626 & YSC 19 Aa,Ab     &  52958.7366 &  $-$11.99 &   0.85 &  $-$0.12 & a \\
00541+6626 & YSC 19 Aa,Ab     &  52984.6649 &  $-$11.36 &   0.85 &   0.08 & a \\
00541+6626 & YSC 19 Aa,Ab     &  53013.6392 &  $-$10.85 &   0.86 &   0.12 & a \\
02249+3039 & HDS 314 Aa,Ab    &  49198.8524 &    1.62 &   0.47 &   0.61 & a \\
02249+3039 & HDS 314 Aa,Ab    &  49237.8599 &    0.18 &   0.42 &  $-$0.85 & a \\
02249+3039 & HDS 314 Aa,Ab    &  49261.6807 &    1.49 &   0.43 &   0.45 & a 
\enddata 
\tablenotetext{a}{Components: a -- primary, b -- secondary, c -- tertiary.
  RVs from the CfA digital speedometers have no instrument codes.  Otherwise, 
RVs measured from the TRES spectra are marked by T,
  RVs from \citet{Sperauskas2019} by S, RVs from \citet{TS02} by t, and RVs from {\it Gaia} DR2 by G.}
\end{deluxetable*}

Individual positional measurements and their residuals from the orbits
are   listed    in   Table~\ref{tab:speckle},   published    in   full
electronically. Similarly,  the RVs used in combined  orbits and their
residuals are listed in  Table~\ref{tab:rv}. These tables also contain
measurements   and  residuals   for  combined   orbits   presented  in
Section~\ref{sec:sb}    and   for    the   hierarchies    covered   in
Sections~\ref{sec:62556}, \ref{sec:65026}, and \ref{sec:85209}.

Below we  comment on some of  those systems. The  information is taken
from    the    MSC    \citep{MSC}    and   the    {\it    Gaia}    DR2
\citep{Gaia}. Comparison  of the  {\it Gaia} short-term  proper motion
(PM)  with the  long-term PM  computed from  the {\it  Gaia}  and {\it
  Hipparcos}  positions  allows  in   some  cases  to  detect  the  PM
difference    $\Delta    \mu$    (astrometric    acceleration),    see
\citet{Brandt2018}.  It should be directed oppositely to the motion of
the  secondary  component predicted  by  our  orbits.  We denote  here
subsystems  by joining  components'  designations with  a comma  (e.g.
A,B).   In a triple  system with  a tertiary  component C,  AB without
comma  means the  center of  gravity of  the inner  pair A,B  and AB,C
refers to the outer orbit.

01500+7456  MLR 297  AB,C (HIP  8533) moves  slowly on  a  200 yr orbit
(Figure~\ref{fig:grade5}). The  0\farcs06 subsystem BAG~16  A,B with a
much shorter  expected period of  $\sim$15 yr was resolved  in 2018.08
but  closed down in  2018.9 (we  use only  the latter  measurement for
the outer orbit).  The {\it Gaia}  parallax of 7.4$\pm$0.7 mas has a large
error  that might  be  caused  by the  subsystem  A,B.  However,  some
measurements attributed in  the WDS to this inner  pair actually refer
to  AB,C and fit its orbit.

02529+5300  A 2906  A,B (HIP  13424,  HR~846) has  passed through  the
periastron  of  its eccentric  ($e=0.9$)  90 yr  orbit  in 1973  and
presently is  again near the apastron,  close to the  position where it
has  been discovered  a century  ago by  R.~Aitken.   The measurements
cover  one full  revolution and  constrain the  orbit  reasonably well
(grade 4).  The tertiary component C at 1\farcs57 has been known since 1825.
Although {\it Gaia}  gives for C slightly discrepant  parallax and PM,
the  outer pair  is likely  physical,  considering the  low chance  of
finding such a bright ($V_{\rm C}=7.25$  mag) star so close to AB. The
estimated period of  AB,C is $\sim$2 kyr and  its observed slow motion does
not contradict this estimate.

03127+7133 STT 50  A,B (HIP 14944).  The orbit  of \citet{Sca2012b} is
corrected here  from $P=309$  yr to $P=583$  yr and assigned  grade 4.
The subsystem Sca~171 Aa,Ab is not  confirmed, so this appears to be a
binary.

04192+6135 BU  1333 B,C.  The measurements cover  one century  and the
orbital period  is about 180  yr.  The pair  is closing and  will pass
through the periastron in 2044. The main component A (HIP 20157) remains at
5\farcs3 separation from BC since the discovery of A,B by W.  Struve in 1830; the
period of A,BC is $\sim$15 kyr.

04480+5645 HDS~617 Aa,Ab is the bright star 4~Cam (HR~1511, HIP~22287)
resolved by {\it  Hipparcos} at 0\farcs6.  Now it  closed to 0\farcs23
and approaches the periastron  of its preliminary eccentric ($e=0.87$)
90 yr  orbit, predicted for  2027.  The faint ($V=13.2$  mag) physical
tertiary component B is located at 13\farcs8 from A.

05237+7347 A 843 A,B (HD~34254) was discovered in 1904 by R.~Aitken at
0\farcs7 and presently has closed  down to 0\farcs04.  Its first orbit
with  $P=182$   yr  is  not  sufficiently   constrained  by  available
observations. It is  fitted with a fixed eccentricity  of $e=0.80$ and
fixed $\omega$, chosen  to yield a reasonable mass  sum.  The physical
tertiary component C is located at 80\arcsec ~from AB.

07211+6740 JNN 55 B,C  is a nearby (26 pc) pair of  M3V and M5V dwarfs
discovered in 2008  \citep{Jnn2012}. Its first orbit with  $P=14$ yr is
computed   from  measurements   at  five   epochs.   The   period  and
eccentricity  are  fixed  to  match  the  expected  mass  sum  of  0.5
\msun.  Further monitoring  for a  few more  years will  constrain all
elements.   The K0V  primary component  A  (HIP 35628)  is located  at
20\farcs7 from BC.

07295+3556  JNN 57  Aa,Ab is  a  young pair  of M1V  dwarfs at  42\,pc
belonging  to the  $\beta$~Pictoris  moving group  \citep{Alonso2015}.
Its first orbit with $P=18.4$ yr is sufficiently well constrained by nine
measurements taken during 11 years.  The tertiary component B is located at
95\farcs5.

08307+4645 YSC  3 Aa,Ab (HIP  41739, distance 118\,pc).  Owing  to the
small $\Delta m$, the quadrants  are not certain; they were changed in
two  instances to match  our circular,  still preliminary  orbit.  The
tertiary component B at 1\farcs2  is also measured here; its estimated
period is $\sim$900 yr.

09186+2944 A 221 Ba,Bb (HD  80101) has been discovered by R.~Aitken in
1901;   its 183 yr orbit presented here  is still tentative.
The   visual   component   A   (HJL   1054  AB)   at   343\arcsec   ~is
optical. However,  the pair  Ba,Bb contains a  spectroscopic subsystem
with a period  of 52.8 days. {\it Gaia} does  not provide the distance
measurement.  The dynamical  parallax  of 14.2\,mas  derived from  the
orbit matches the photometric distance.

09354+3958 COU 2084 Aa,Ab (HIP~47053).   This is a quintuple system at
79 pc distance  from the Sun.   The outer 25\arcsec  ~pair A,B has  an estimated
period  of  40   kyr.   The  component  B  (HIP~47054)   is  a  28 day
spectroscopic   binary,  and  the   component  Aa   is  also   a  tight
spectroscopic and eclipsing pair with $P=1.07$ d and the primary of
F2V  spectral  type.   Our orbit  of  Aa,Ab  with  $P=50$ yr  is  well
constrained.  Note that  the {\it Gaia} DR2 parallaxes of  A and B are
formally    discordant   (12.63$\pm$0.39   and    14.61$\pm$0.29   mas,
respectively), possibly being biased by the subsystems.

10454+3831 HO 532 A,C (HIP  52600). The orbit by \citet{Mnt2000a} with
$P=161$   yr,  revised   here   to  $P=326$   yr,   is  still   poorly
constrained. This is a nearby (13.6  pc) K7V dwarf.  We do not confirm
the 0\farcs4 subsystem CHR 191 A,B and believe that it is spurious. It
has been  resolved three  times in  the period from  1983 to  1991, at
similar positions \citep{Hrt1994}, but subsequent speckle observations
detected  only the  wide pair  A,C, now  at 0\farcs7  separation.  The
outer  eccentricity $e=0.90$  is  incompatible with  a 0\farcs4  inner
subsystem,  violating dynamical  stability.   Orbits  with smaller
  eccentricities or larger  inclinations correspond to unrealistically
  small mass sum, so these parameters were fixed.  

11017+3641 HDS 1574 Aa,Ab (HIP 53903). The new orbit with $P=47$ yr is
relatively  well  defined.  Comparison  between  {\it  Gaia} and  {\it
  Hipparcos} reveals $\Delta \mu  = (-1.3, -17.0)$ mas~yr$^{-1}$ which
matches  the orbital  motion of  Aa,Ab.   The companion  B located  at
46\farcs8  from A  is also  a close  pair COU~1422  Ba,Bb  with direct
revolution  (Aa,Ab is  retrograde).   Therefore, this  is a  quadruple
system of 2+2 hierarchy with non-coplanar inner subsystems.  We compute
a preliminary orbit of Ba,Bb with $P=173$ yr.

11151+3735  CHR 192  Aa,Ab (HIP  54941) has  a preliminary  orbit with
$P=200$  yr  (Figure~\ref{fig:grade5}).   The  tertiary  component  at
0\farcs6 has been  measured as well. This is  a distant ($\sim$300 pc)
triple  system with  a G5  giant primary  component.  We consistently
measure smaller $\Delta m$  at shorter wavelengths, indicating that Ab
is  hotter  than  Aa.    {\it  Gaia}  measured  a  negative  parallax,
apparently being confused by the multiple nature of this source.

12089+2147 HDS 1714 Aa,Ab (HIP 59233) has moved almost on a straight line
since its  resolution by {\it  Hipparcos} in 1991.25.  The  absence of
significant $\Delta  \mu$ supports the long period  of our preliminary
orbit.  The physical tertiary component B is at 15\farcs3.

14353+4302 LSC 56 Aa,Ab (HIP 71366). The first tentative edge-on orbit
with  $P=11$ yr  is determined  here.  The  separation does  not exceed
50\,mas.   We  also  measured  the  third star  B  at  0\farcs5.   The
estimated period of A,B is $\sim$400 yr. 

16147+3352 YSC 152  Ea,Eb (HIP 79551) belongs to  the nearby (22.7 pc)
quintuple  system.   The  primary  component  (HIP  79607,  GJ  615.2,
$\sigma$~CrB) is a  726 yr visual binary A,B with  a 1.1 d subsystem
Aa,Ab that  has been resolved interferometrically. The  pair Ea,Eb, at
634\arcsec  ~from  AB,  was   known  as  a  52 yr  astrometric  binary
\citep{Hei1990d}.   Here we  determine  the elements  of its  circular
visual orbit with  the same period. So far,  the measurements of Ea,Eb
cover only 10  years (owing to the large $\Delta  V \approx 3.3$ mag,
this  pair has  not  been resolved  by  {\it Hipparcos}).   Comparison
between  the {\it Gaia}  and {\it  Hipparcos} astrometry   reveals
$\Delta \mu$ of  the star E that matches its  orbital motion and estimated
mass ratio.

19514+4044  COU  2530  A,B  (HIP  97706)  is  a  quadruple  system  at
81~pc.  The outer  4\farcs4 pair  AB,C (ORL  6) is  confirmed  by {\it
  Gaia}.  We compute  here the first visual orbit  of the intermediate
subsystem A,B with  $P=55$ yr.  Double lines in  the spectrum reported
by  \citet{Guillout2009} suggest  an  inner subsystem  Aa,Ab, but  its
orbit is unknown.

20599+4016 HDS~2989 Da,Db (HIP 103052, K7Ve, 41pc).  We determined the
preliminary  orbit of  Da,Db  with $P=213$  yr.  It has  a common  PM,
parallax and  RV with HIP~103641,  located at projected  separation of
20\arcmin  (0.24 pc)  from  D. The  component  A itself  is triple  (a
2\farcs2    visual    binary     with    a    112 d    spectroscopic
subsystem). Although   A and D are obviously related to each other, it
is not clear  at the moment whether they  are gravitationally bound or
are just members of some unknown moving group. 


21308+4827 A 770 A,B. We computed a tentative edge-on orbit with
$P=146$ yr. The outer companion at 1\farcs1 is also measured; its
estimated period is $\sim$1 kyr. 

22139+3943. We computed a tentative orbit of McA 70 Aa,Ac (HIP 109754,
K3III) and have not resolved the 0\farcs09 subsystem Aa,Ab. We 
question the existence of Aa,Ab because it has not been confirmed by
numerous speckle observations of Aa,Ac. 

22359+3938  CHR 112 Aa,Ab  (HIP 111546,  B1Ve) is  a distant  (500 pc)
pair.   We compute  its first  orbit with  $P=42$ yr,  reasonably well
constrained (grade 3).  The visual  components B at 22\farcs3 and D at
81\farcs6 have similar  {\it Gaia} parallaxes and could  be members of
the same cluster, rather than components of a hierarchical system.

\subsection{Combined visual-spectroscopic orbits}
\label{sec:sb}


\begin{deluxetable*}{l l cccc ccc ccc}    
\tabletypesize{\scriptsize}     
\tablecaption{Combined visual-spectroscopic orbits
\label{tab:comborb}          }
\tablewidth{0pt}                                   
\tablehead{                                                                     
\colhead{WDS} & 
\colhead{Disc.} & 
\colhead{$P$} & 
\colhead{$T$} & 
\colhead{$e$} & 
\colhead{$a$} & 
\colhead{$\Omega_A$ } & 
\colhead{$\omega_A$ } & 
\colhead{$i$ } & 
\colhead{$K_1$}  &
\colhead{$K_2$}  &
\colhead{$\gamma$} 
\\
\colhead{\it HIP } &
& 
\colhead{(yr)} &
\colhead{(yr)} & &
\colhead{(arcsec)} & 
\colhead{(deg)} & 
\colhead{(deg)} & 
\colhead{(deg)} &  
\multicolumn{3}{c}{ (km~s$^{-1}$) } 
}
\startdata
00541$+$6626 & YSC 19 Aa,Ab &10.868& 2013.19  & 0.185 & 0.0718 & 140.4 & 195.2 & 111.4  & 8.77 & \ldots & $-$9.60  \\
{\it 4239}      & & $\pm$0.060& $\pm$0.33  & $\pm$0.016 & $\pm$0.0025 & $\pm$2.6 & $\pm$12.0 & $\pm$2.1  & $\pm$0.55&\ldots  & $\pm$0.38  \\
02249$+$3039 & HDS 314 A,B & 43.2 & 2026.5  & 0.89  & 0.271   & 214.8 & 110.3 & 133.5  & 6.29 & \ldots &  2.25  \\
{\it 11253}      & & $\pm$2.7& $\pm$2.5  & fixed & $\pm$0.070  & $\pm$31.2& $\pm$12.4  & $\pm$24.2 & $\pm$1.44 & \ldots  & $\pm$0.63  \\
03400$+$6352 & HU 1062 B,C & 11.42 & 2000.601 & 0.475 & 0.146 & 232.4 & 157.8 & 70.0       & 7.22 & 9.95   & 16.76 \\
{\it 17126}   & & $\pm$0.08 & $\pm$0.055 & $\pm$0.016 & $\pm$0.002 & $\pm$1.1 & $\pm$2.9 & fixed & $\pm$0.14&$\pm$0.36 & $\pm$0.10    \\
03562$+$5939 & HDS 497 A,D & 28.87 & 2006.630 & 0.456 & 0.442 & 246.7 & 264.6 & 51.2       & 4.04 & \ldots & $-$18.37 \\
{\it 18413}  & & $\pm$0.24 & $\pm$0.143 & $\pm$0.009 & $\pm$0.048 & $\pm$1.6 & $\pm$2.8 & $\pm$5.0& $\pm$0.08& \ldots & $\pm$0.05    \\
05465$+$7437 & YSC 148 A,B & 11.140 & 1994.922 & 0.347 & 0.1638 & 172.1 & 323.4 & 43.0 & 4.45 & \ldots  & 17.07 \\
{\it 27246}  &    & $\pm$0.039 & $\pm$0.068 & $\pm$0.008 & $\pm$0.0023 & $\pm$1.8 & $\pm$1.9 & $\pm$1.9& $\pm$0.06  & \ldots &  $\pm$0.04 \\
12490$+$6607 & DEL 4 Aa,Ab & 0.14810& 1993.3055 & 0.080 & \ldots & \ldots    & 135.1 & \ldots & 21.80 & 22.36  & \ldots \\
{\it 62556}  &  & $\pm$0.00001 & $\pm$0.0012 & $\pm$0.004 & \ldots  & \ldots & $\pm$3.0 & \ldots & $\pm$0.20  &  $\pm$0.22 & \ldots \\
12490$+$6607 & DEL 4 A,B & 7.370  & 1994.490 & 0.511 & 0.356        & 201.5  & 263.4 & 78.1     & 5.22 & 10.80  & $-$10.67 \\
             &    & $\pm$0.013&$\pm$0.022& $\pm$0.012 &$\pm$0.007&$\pm$0.7&$\pm$0.7& $\pm$0.7& $\pm$0.14  & $\pm$0.31 &  $\pm$0.06 \\
12576+3514 & BWL 53 Aa,Ab & 4.595 & 2010.26 & 0.184 & 0.133 & 352.4   & 314.6   & 134.7 & 4.39 & \ldots  & $-$6.90 \\
{\it 63253}  &    & $\pm$0.096 & $\pm$0.10   &$\pm$0.015 & $\pm$0.005 &$\pm$2.4 &$\pm$7.0 &$\pm$3.9& $\pm$0.15  & \ldots &  $\pm$0.08 \\
13198+4747 & HU 644 A,B & 49.077& 2017.400& 0.221 & 1.551 & 271.0   & 253.1   & 94.31 & 1.07 & \ldots  &    0.96 \\
{\it 65026 }   &    & $\pm$0.07  & $\pm$0.45 &$\pm$0.004 & $\pm$0.005 &$\pm$0.1 &$\pm$0.3 &$\pm$0.05 & $\pm$0.10  & \ldots &  $\pm$0.06 \\
13198+4747 & CHR 193 Aa,Ab & 1.22765 &2014.909 & 0.067 & 0.1115 & 279.8   & 356.0   & 87.7 & 8.703 & \ldots  &   \ldots \\
       &    &$\pm$0.00016  & $\pm$0.022 &$\pm$0.006 & $\pm$0.0020 &$\pm$0.9 &$\pm$6.5 &$\pm$1.0 & $\pm$0.094  & \ldots &  \ldots \\
17247+3802 & HSL 1 Aa,Ab & 1.22496 & 1986.381 & 0.160 & 0.0290 & 53.2   & 358.3   & 81.5 & 16.30 & 17.44 &   \ldots \\
{\it 85209}   &    &$\pm$0.00010  & $\pm$0.002 &$\pm$0.002 & $\pm$0.0009 &$\pm$1.7 &$\pm$0.9 &$\pm$3.3 & $\pm$0.11  & $\pm$0.13 &  \ldots \\
17247+3802 & HSL 1 Aab,Ac & 34.17   & 2011.39  & 0.148 & 0.306  & 57.8   & 277.2   & 92.7 & 3.62  & 7.32  &  0.18 \\
       &    &$\pm$0.48     & $\pm$0.72  &$\pm$0.011 & $\pm$0.019  &$\pm$0.2 &$\pm$1.1 &$\pm$0.2 &$\pm$0.20  & $\pm$0.78 & $\pm$0.04 \\
17422$+$3804 & RBR 29  Aa,Ab & 6.114 & 2008.475 & 0.260 & 0.1033 & 203.1       & 53.7    & 105.8 & 8.61  & 11.25  & $-$44.25 \\
{\it 86642} &    &  $\pm$0.009 & $\pm$0.015         & $\pm$0.004 & $\pm$0.017 &$\pm$5.9&$\pm$1.0 &$\pm$3.4&$\pm$0.03&$\pm$0.07&$\pm$0.03 
\enddata 
\end{deluxetable*}

\begin{figure}
\epsscale{1.1}
\plotone{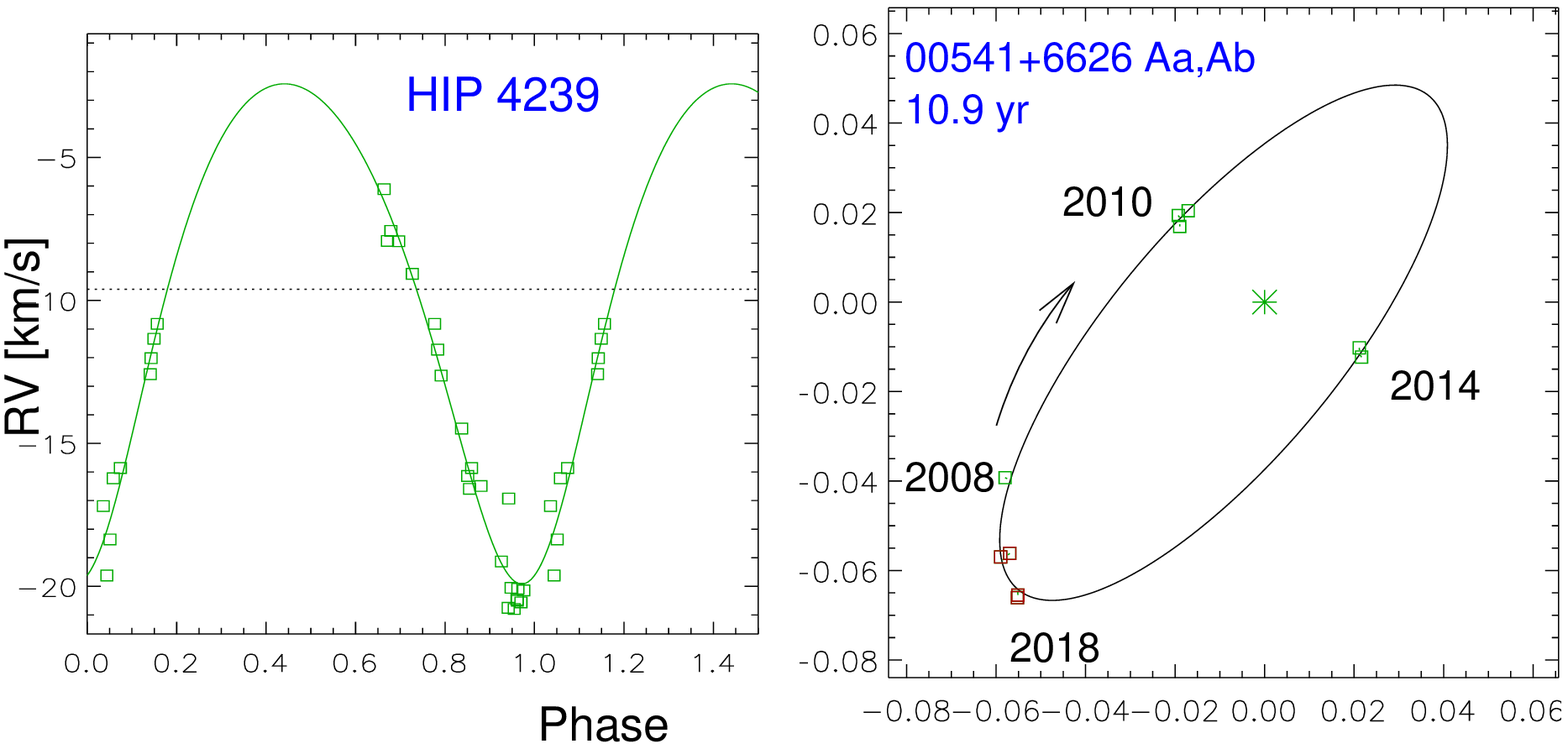}
\plotone{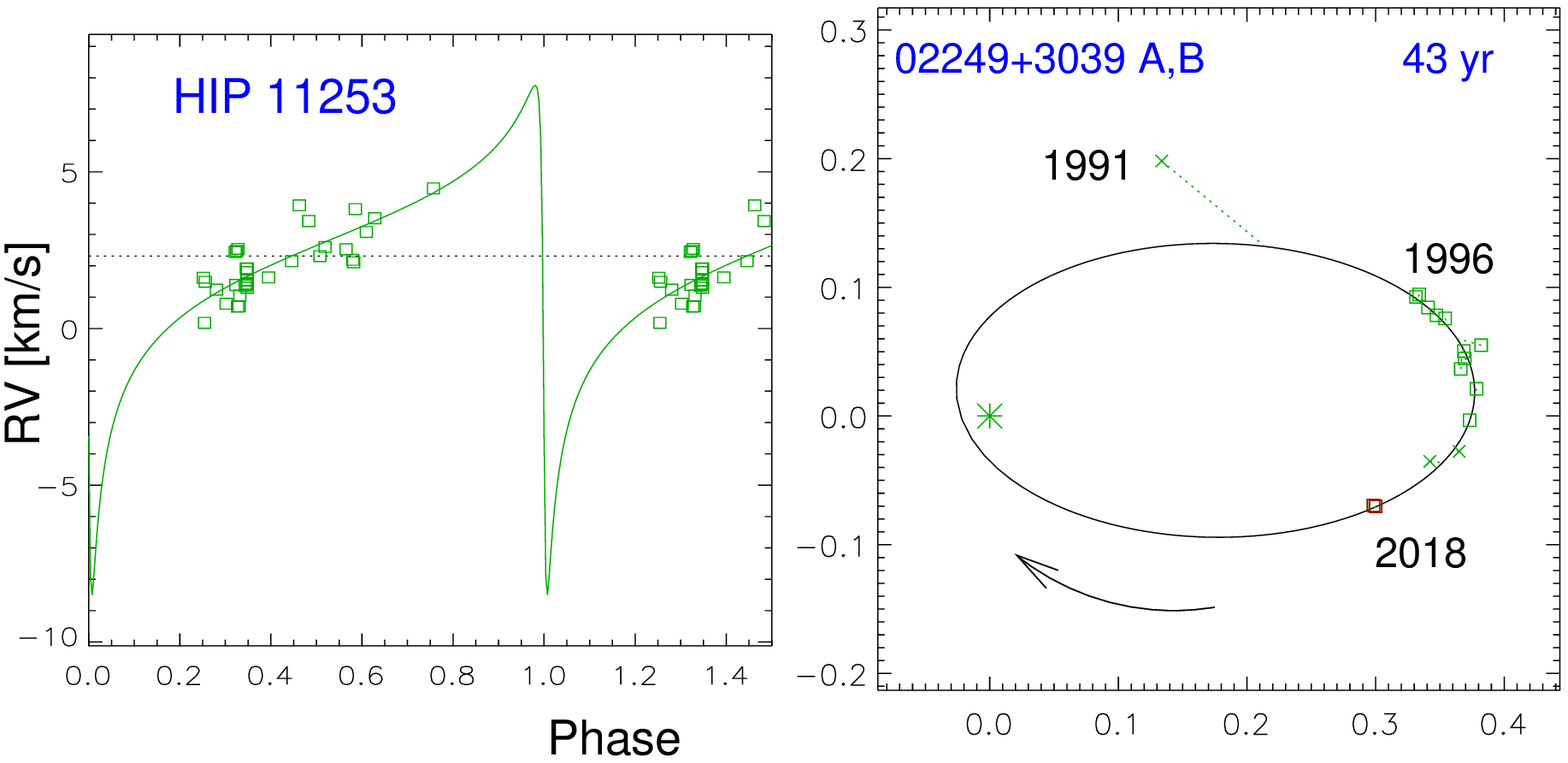}
\caption{Combined spectro-interferometric orbits of HIP~4239 (top)
  and HIP~11253  (bottom). In this and following plots, the left
  panel shows the RV curve (green line and squares for the primary
  component, blue line and triangles for the secondary component). The
  right-hand panel shows orbits in the plane of the sky.
\label{fig:comb1} }
\end{figure}

\begin{figure}
\epsscale{1.1}
\plotone{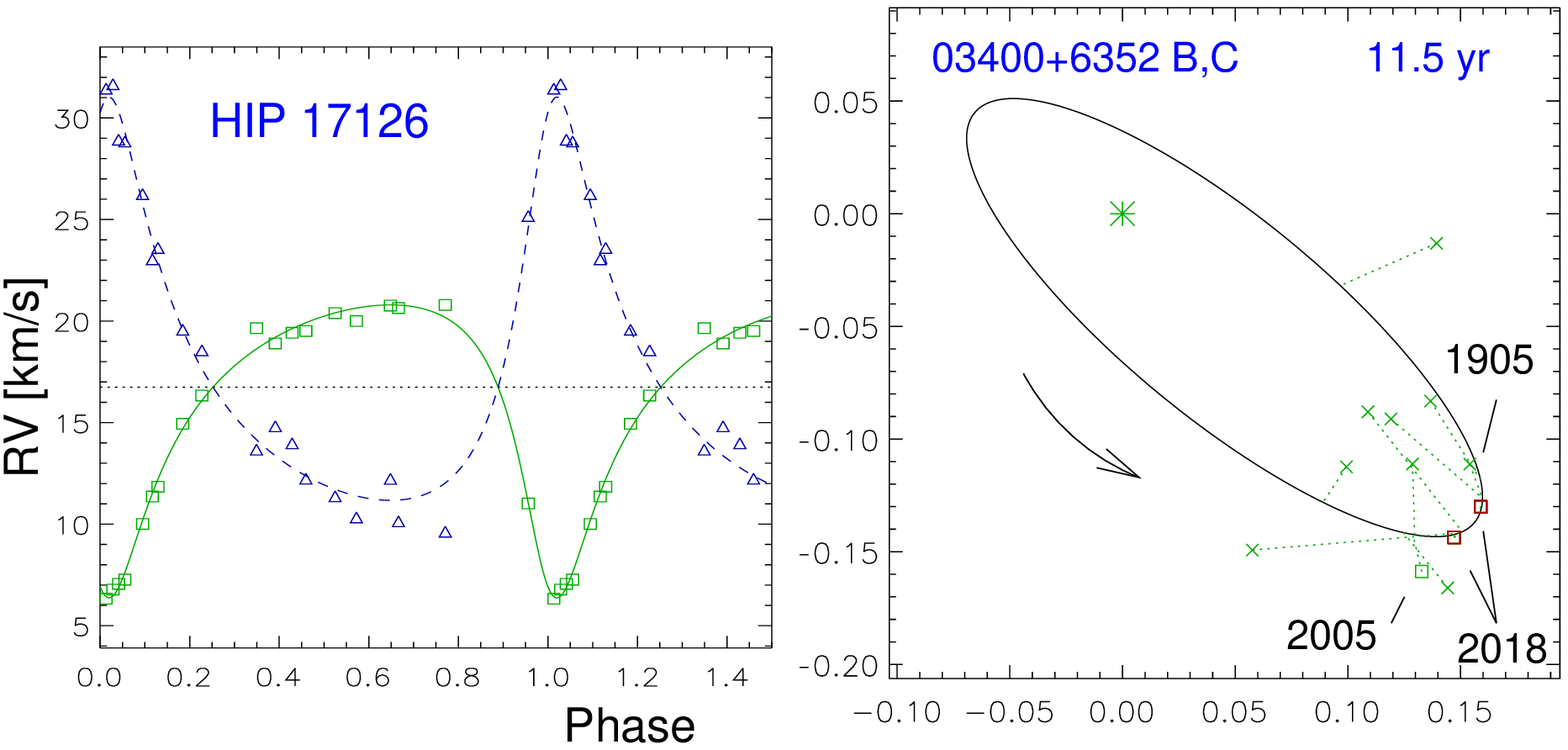}
\plotone{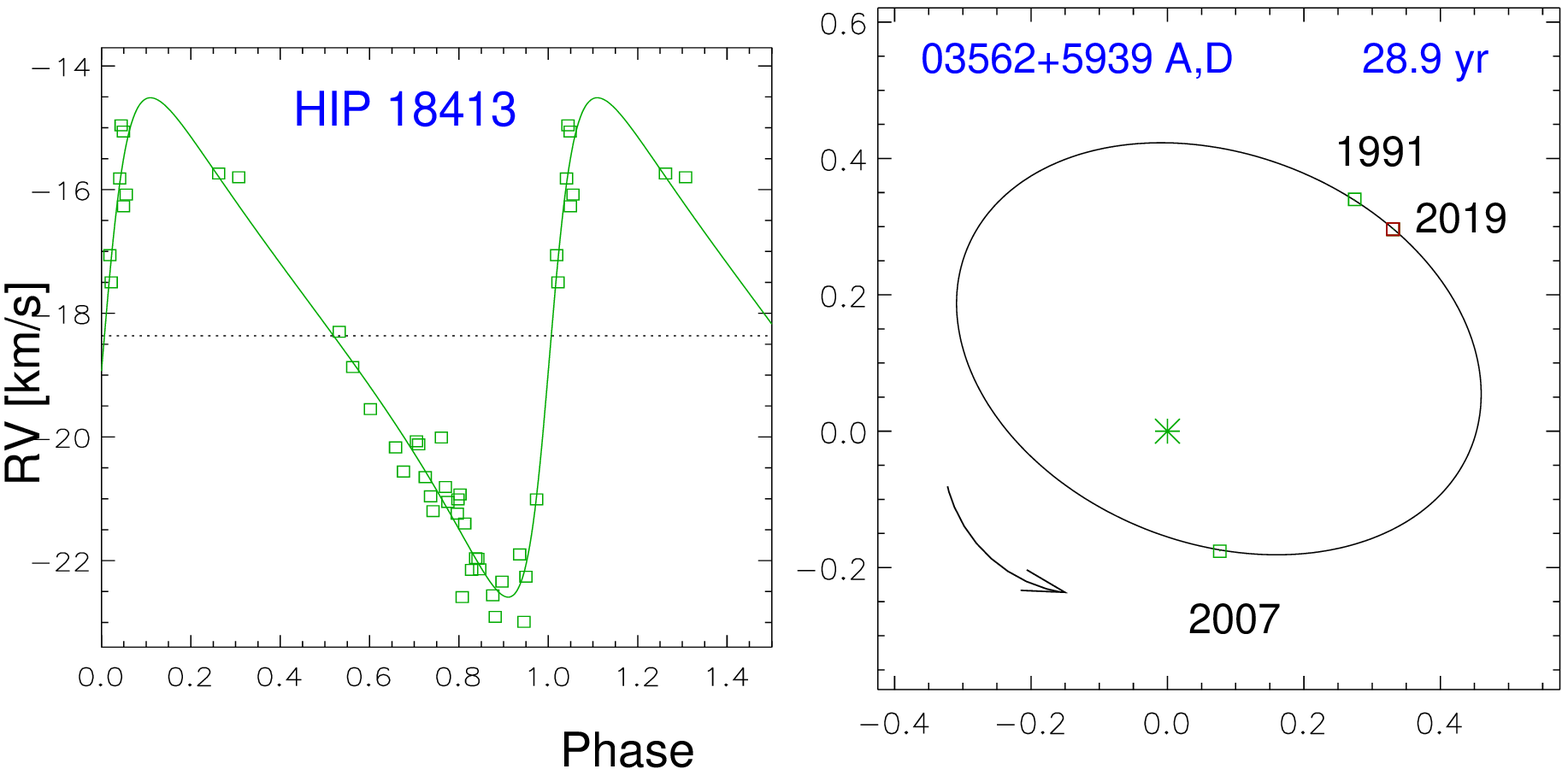}
\caption{Combined spectro-interferometric orbit of HIP~17126 (top) and
  HIP~18413 (bottom).
\label{fig:comb2} }
\end{figure}

\begin{figure}
\epsscale{1.1}
\plotone{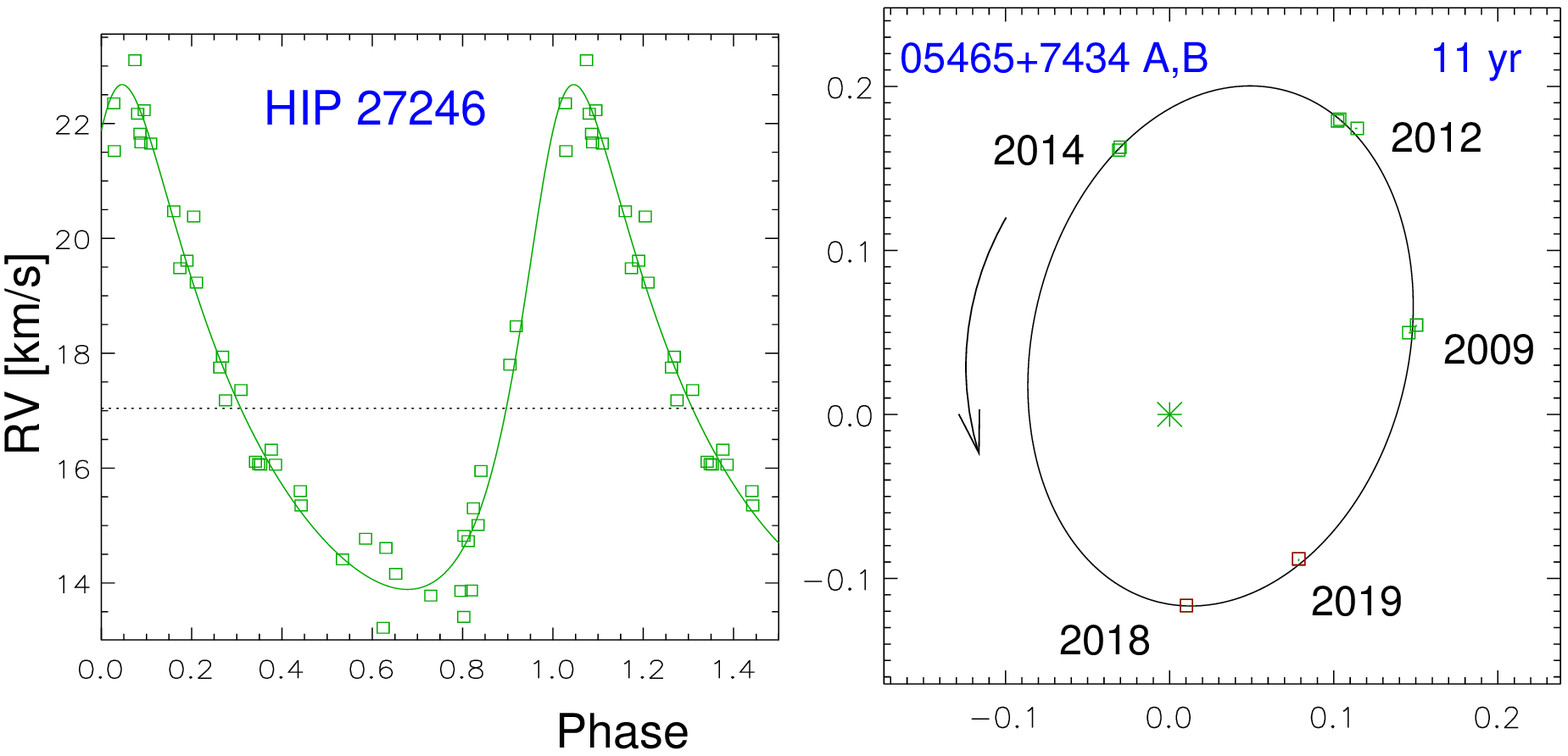}
\plotone{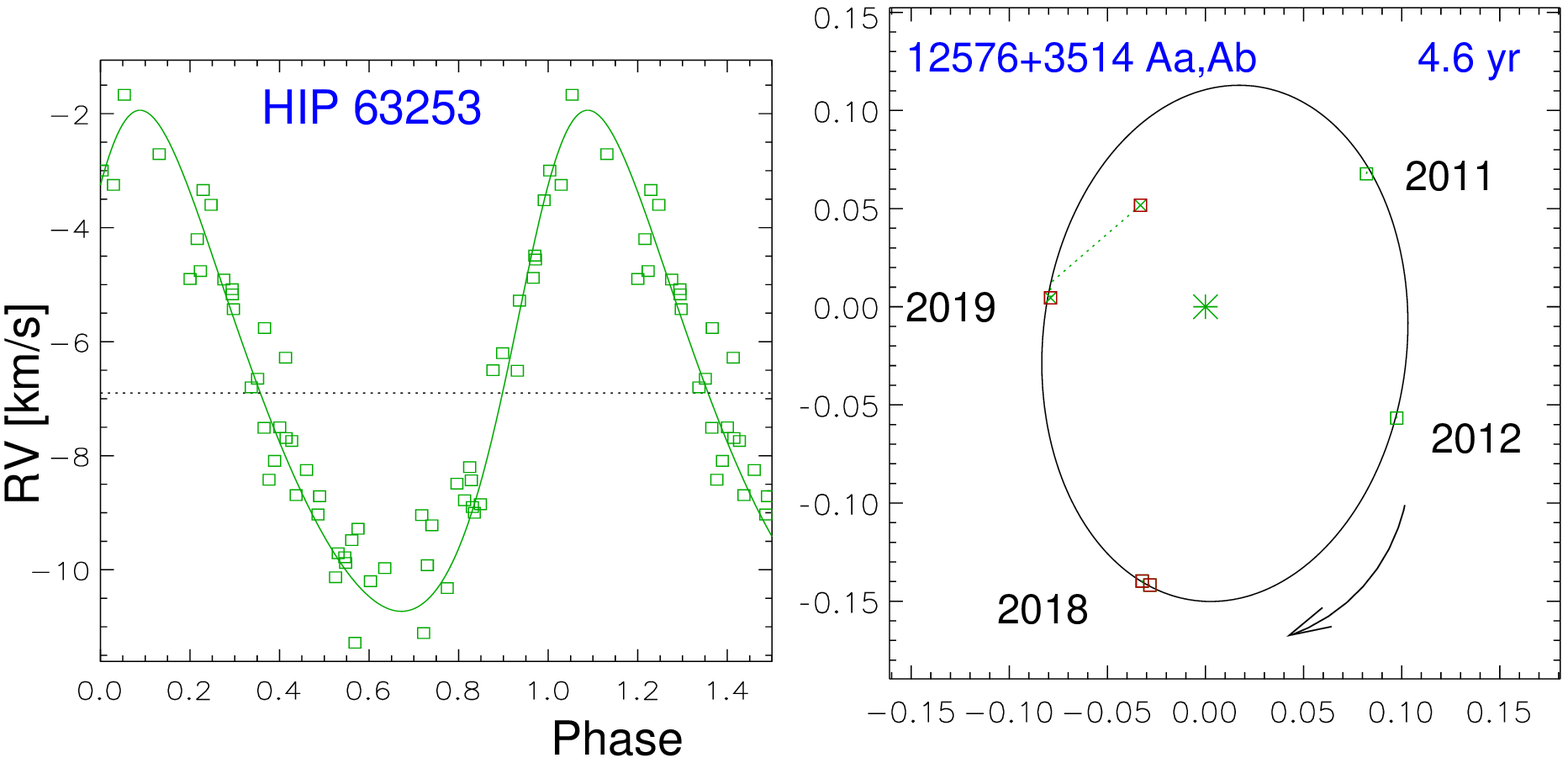}
\caption{Combined spectro-interferometric orbits of HIP~27246 (top)
  and HIP~63253 (bottom).
\label{fig:comb3} }
\end{figure}

\begin{figure}
\epsscale{1.1}
\plotone{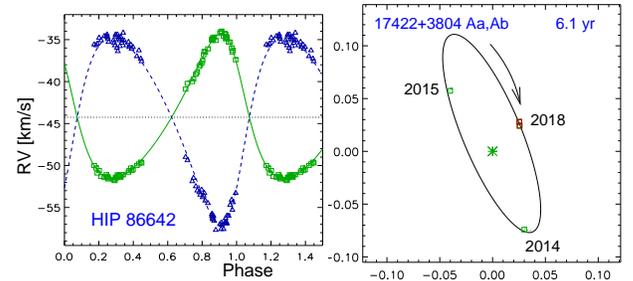}
\caption{Combined  spectro-interferometric orbit  of   HIP~86642.
\label{fig:comb4} }
\end{figure}

In  this  Section,  we  present  orbits computed  by  fitting  jointly
positional measurements  and radial velocities  (RVs).  Their elements
are given in Table~\ref{tab:comborb}. The  RV curves and orbits in the
plane   of   the   sky   are   plotted   in   Figures~\ref{fig:comb1} to
 \ref{fig:comb4}.   Each system is commented below
except HIP~62556, 65026, and 85209 discussed in the following Sections.

00541+6626  YSC 19  Aa,Ab  (HIP  4239).  Its  first  orbit is  defined
independently by both positional measurements and the CfA RVs covering
a  time span of  16 yrs.   The combined  elements are  quite accurate,
$P=10.87 \pm  0.06$ yr  (Figure~\ref{fig:comb1}, top). The  {\it Gaia}
parallax  of 9.49$\pm$0.05\,mas  corresponds to  the mass  sum  of 3.7
\msun, in rough agreement with the  expected mass of two F0V stars (Aa
and Ab  are similar, $\Delta  m = 0.4$  mag).  Lines of  the secondary
component Ab are  detectable, but their RVs do  not match the expected
RVs in  the 11-yr orbit,  so Ab may  contain a subsystem. We  use here
only  the RVs  of Aa.   The  {\it Gaia}  $\Delta \mu$  is small,  only
$\sim$0.1 of the orbital  motion, with opposite direction as expected.
The  outer physical  component  B  at 0\farcs9  moves  slowly with  an
estimated period of $\sim$500 yr.

02249+3039 HDS~314  Aa,Ab (HIP  11253, HD~14784). The orbit  of \citet{Lin2012b}
with $P=82.18$  yr does not  agree well with recent  measurements and,
moreover, yields  an unrealistically  small mass sum.   Its adjustment
using  only  relative  astrometry   does  not  help;  the  first  {\it
  Hipparcos}   resolution  obviously   deviates  from   the  remaining
measurements. However,  use of  the RVs improves  the situation,
although they  do not cover the  full period.  To the  34 RVs measured
with the CfA  Digital Speedometers between 1993 and  2009 we added one
more from  TRES obtained in 2009, as  well as the mean  RV reported by
{\it Gaia}.  The resulting  orbit (Figure~\ref{fig:comb1}, bottom) has a large
eccentricity of $e=0.89$ and  is still not fully constrained, although
the mass  sum is  now realistic.  Further  monitoring of this  pair by
both speckle interferometry and  RVs as it approcaches the periastron,
predicted for 2026, will eventually lead to a reliable orbit.  This is
a quadruple  system with a 2+2 hierarchy at  55\,pc from the  Sun  (the
component C, at 20\farcs4 from A, is itself a 0\farcs6 pair).

03400+6352  HU 1062 B,C  (HIP 17126)  forms a  physical pair  with the
component A (HIP~17118), at  45\farcs7 distance from each other.  Both
stars have  common parallaxes  (23.27$\pm$0.030\,mas for A),  PMs, and
RVs.   The period  of A,B  is $\sim$50\,kyr.   The pair  B,C  has been
resolved in 1905 at  0\farcs19, but its visual micrometer measurements
accumulated during the  past century were too rare  and discordant for
orbit  determination.  Prior  to our  observations at  WIYN,  only one
speckle measurement has been made in  2005. On the  other hand, the RV
monitoring at CfA  revealed B,C as a double-lined  binary with a period
of $\sim$11 yr.  The combined orbit in Figure~\ref{fig:comb2}, top,
was computed by fixing the  inclination to $i=70\degr$ to bring the
RV amplitudes in  agreement with estimated masses of B  and C (1.0 and
0.8 \msun,  respectively) and with the  mass sum of  1.87 \msun deduced
from the  parallax.  Further  high-resolution observations of  B,C are
needed  to cover  the full  orbit and  constrain its  inclination. The
$\Delta \mu$ of the unresolved star  BC measured by {\it Gaia} and {\it
  Hipparcos}, $(+8.1,  +2.6)$ \masyr, is opposite to  the orbital speed
of C relative to B, $(-48, -31)$ \masyr in 2015.5.

03562+5939 (HIP  18413, HD~24409,  GJ~3257) is a  pair of dwarfs  at a
distance of  22.7 pc from the Sun;  it belongs to the  25-pc sample of
solar-type stars.  The pair A,D  was first resolved by {\it Hipparcos}
(HDS 497  A,D).  The  RVs were  monitored by D.~L.   using the  CfA RV
spectrometers from  1993.1 to 2008.2,  a total of 36  measurements. We
also used one RV from TRES and  the mean RV measured by {\it Gaia}.  A
preliminary  period  of 28  yrs  has  been  determined from  the  RVs.
Positional measurements  are available only at  three epochs: 1991.25,
2007.6,   and   2018.9.    Nevertheless,   the   combined   orbit   in
Figure~\ref{fig:comb2},  bottom,  is defined  reasonably  well if  the
quadrant in 2007.6 is changed.  The rms residuals in RV are 0.41 \kms.
At the  {\it Gaia} epoch of  2015.5, the companion moved  on the orbit
with a speed of $(+18, +49)$  \masyr, two times faster and in opposite
direction compared to $\Delta \mu = (-10.6, -25.0)$ \masyr measured by
{\it Gaia} and {\it Hipparcos}.  This confirms the orbit and indicates
comparable masses of A and D, despite substantial magnitude difference
($\Delta  m =  3.4$ mag  in  the $V$  band).  Lines  of the  secondary
component D are not detected in the spectra, but they could reduce the
RV amplitude  by blending with the  lines of A.  The  RV amplitude and
the estimated mass of A, 1.0  \msun, correspond to the minimum mass of
0.64 \msun for D.  The  faint tertiary companion E at 9\farcs5 (BUP~48
A,E) is  physical.  The distant companions  B and C listed  in the WDS
are optical.

05465+7437 (YSC 128  A,B; HIP 27246, distance 39  pc) was resolved for
the first time  in 2009.75 by \citet{Hor2012a} at  0\farcs15.  Its RVs
measured from 1987.3 to 1999.9 are published by \citet{Latham2002} who
also derived the spectroscopic orbit;  we do not copy these RVs in
  Table~\ref{tab:rv}.  By combining  RVs with positional measurements
at  six  epochs  (including  two  from this  program),  we  derive  an
excellent combined  orbit (Figure~\ref{fig:comb3}, top)  with a period
of 11.14$\pm$0.05  yr.  An astrometric  orbit with similar  period has
been  published  by  \citet{Jnc2005}.   The tertiary  companion  C  is
located at 10\farcs9.  Its  parallax of 26.80$\pm$0.04 mas is measured
by {\it Gaia}  more accurately than the parallax of  AB.  The mass sum
derived from the orbit and the  parallax is 1.82 \msun; it matches the
masses  of A  and B  estimated  from the  luminosity, as  well as  the
secondary  mass computed  from  the RV  amplitude  and inclination. 

12576+3514 (HIP  63253, GJ  490) is a  low-mass quadruple system  of M
dwarfs at 20  pc from the Sun. The outer  16\arcsec ~pair LDS~5764 A,B
has been known since  1950.  Another star C with a  similar PM, at 12\farcm7
from  A,   is  unrelated  according   to  its  {\it   Gaia}  parallax.
\citet{Bwl2015} resolved in 2011 stars  A and B into tight pairs Aa,Ab
and Ba,Bb  with separations of 0\farcs13  and 0\farcs17, respectively.
We measured both subsystems in  2018 and re-observed them in 2019. The
observations in 2019 were repeated  twice on the same night because on
the first attempt the results  were affected by telescope vibration. We
obtained different position angles of Aa,Ab on this night and consider
both our measurements uncertain,  given the small 0\farcs06 separation
and  the  large  $\Delta  m  \sim  3$ mag.  On  the  other  hand,  the
measurements of Ba,Bb on that night are mutually consistent.

A spectroscopic orbit of Aa,Ab with  a period of 4.6 yr was derived by
one  of  us  (G.  T.)  from  45 RVs  measured  with  the  CfA  Digital
Speedometers between 1984  and 2008.  We use these RVs  and the 11 RVs
from   \citet{Sperauskas2019}    to   derive   the    combined   orbit
(Figure~\ref{fig:comb3}, bottom).  The weighted rms RV residuals are 0.60
\kms.  Despite the short period,  the RV amplitude is moderate, $K_1 =
4.4$  \kms, because  the orbit  has an  inclination of  135\degr.  Our
discordant measurements  in 2019 are  assigned a very low  weight.  We
adopt the {\it Gaia} parallax  of the component B, 49.57$\pm$0.13 mas,
as the  distance   to   the   system   because   the   parallax   of   A,
46.83$\pm$0.28\,mas, is less  accurate. The mass sum of  Aa,Ab is then
0.90  \msun.  Our relative  photometry, $\Delta  V_{\rm Aa,Ab}  = 3.2$
mag, is consistent with $\Delta  K_{\rm Aa,Ab} = 1.69$ mag measured by
\citet{Bwl2015}.  The absolute magnitudes of Aa and  Ab correspond to
dwarfs with masses of 0.60 and  0.30 \msun, matching the orbit. The RV
amplitude  and inclination also  match these  masses.  The  {\it Gaia}
astrometry  supports our  orbit of  Aa,Ab by  measuring $\Delta  \mu =
(32.6, 10.4)$ \masyr.  The computed  speed of orbital motion in 2015.5
is  $(-125.5, -63.8)$  \masyr. The  ratio of  those speeds  equals the
wobble  factor $f  \approx  0.24$, in  agreement  with the  calculated
$f=0.27$.

The pair  Ba,Bb is fainter  than A, $V_{\rm  Ba,Bb} = 13.2$  mag.  The
separation corresponds  to a period  of $\sim$10\,yr; we  observed the
motion of  Ba,Bb  pair by  24\degr ~in one  year. The  orbit of Ba,Bb  can be
determined with a few more measurements. We also derived 45 RVs of the
star  B  from  the  CfA  and  TRES  spectra  taken  between  1984  and
2015. Owing to the faintness of the star B, the errors are large, from
3 to 5 \kms, and we do not  resolve the double lines; the mean RV of B
is $-5.8$ \kms.  As no orbit of Ba,Bb could be  derived yet from these
RVs, they are not presented here.

17422+3804  (HIP 86642,  HD~161613, distance  42 pc)  is  a solar-type
triple system. The double-lined spectroscopic and astrometric orbit of
the  inner  pair  RBR~29  Aa,Ab  with $P=6.1$  yr  has  been  recently
published   by  \citet{Fek2018};    we  do   not  copy   these RVs  in
  Table~\ref{tab:rv}.  Measurements in the plane of the sky in 2013.8
and 2015.5  are available from \citet{Rbr2015d}. We  resolved the pair
in 2018.47  but did not  resolve it in  2018.65.  The elements  of our
combined  orbit (Figure~\ref{fig:comb4})  do not  differ substantially
from  the  published orbit  and,  therefore,  confirm its  astrometric
elements.  The RV amplitudes and  inclination imply the masses of 1.02
and  0.79 \msun  which match  the masses  estimated from  the absolute
magnitudes.  The orbital parallax of 25.3\,mas is to be preferred over
the {\it  Gaia} parallax of 23.94$\pm$0.11\,mas because  the latter is
likely  biased by  the  binary.   The ratio  of  astrometric and  full
semimajor axes,  $f = 0.293$,  implies the mass  ratio $q =  f/(1-f) =
0.42$ if the secondary component Ab has a large $\Delta m$.  The outer
2\farcs2 pair A,B has an estimated period of $\sim$600 yr.

\subsection{HIP 62556, a triplet of M3V dwarfs}
\label{sec:62556}

\begin{figure}
\epsscale{1.1}
\plotone{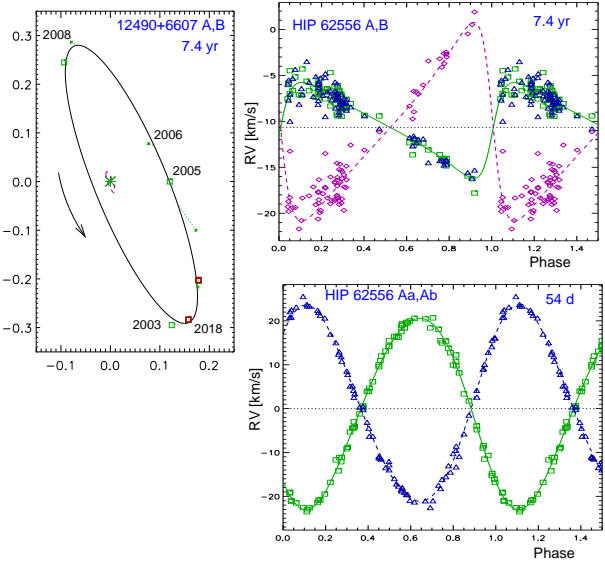}
\caption{Orbits of the inner and outer pairs in HIP~62556 (GJ 487).
\label{fig:62556} }
\end{figure}

This triple  star, located  at 10\,pc from  the Sun, is  designated as
HIP~62556,   GJ~487,   WDS  J12490+6607,   and   BP~Dra.   The   inner
spectroscopic  subsystem  Aa,Ab  with  $P=54$~d has  been  studied  by
\citet{Delfosse1999}.  The estimated semimajor axis of 23 mas makes it
resolvable at  8-m class  telescopes or with  the CHARA array  but, so
far, it  has never  been resolved by  speckle. However,  these authors
resolved in 1997  the outer subsystem A,B (DEL~4)  at 0\farcs23.  They
noted that the  spectrum is triple-lined and that  the outer period is
$\sim$3000 days.   Available astrometry  and two measurements  at WIYN
define the outer visual orbit with $P=7.4$ yr.

This system  has been  observed with the  CfA spectrometers  79 times,
from 1984  to 2003. These data cover  2.6 outer periods and  allow us to
determine the  masses and distance  from the combined outer  orbit. We
used  the {\tt  TRICOR}  software and  the  spectrum of  GJ~725B as  a
template  to derive  the  RVs of  three  stars Aa,  Ab,  and B.  Their
relative fluxes at 5187\,\AA ~are 1:0.97:0.73. The inner subsystem is not
resolved and does  not cause wobble in the outer  orbit because Aa and
Ab   are   nearly   equal.    The   combined   orbit   is   shown   in
Figure~\ref{fig:62556}. The rms RV residuals are 0.68, 0.91, and 1.39
\kms for Aa, Ab, and B, respectively.

As the outer inclination is known,  we determine the masses of A and B
as 0.52  and 0.25 \msun. The  inner mass ratio $q_{\rm  Aa,Ab} = 0.97$
defines  the masses  of  Aa and  Ab,  0.26 \msun  ~each.  The inner  RV
amplitudes then correspond to the inner inclination of 77\degr, so the
orbits are, most  likely, coplanar. The mass sum  and outer semimajor
axis  imply an orbital  parallax  of  103\,mas.  The  {\it
  Hipparcos} parallax is 97.9$\pm$1.8\,mas,  while {\it Gaia} does not
yet provide astrometry of this complex and fast triple system.

\subsection{The low-mass triple system HIP 65026}
\label{sec:65026}

\begin{figure}
\epsscale{1.1}
\plotone{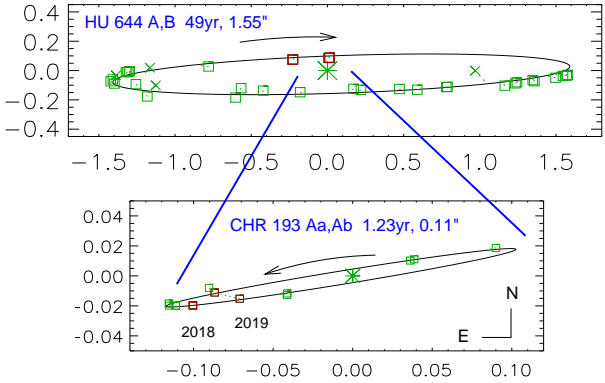}
\caption{Orbits    of   HIP    65026   (WDS    J13198+4747)   computed
  independently. The  upper panel shows  the outer system HU  644 A,B;
  the lower panel shows the inner subsystem CHR~193 Aa,Ab.
\label{fig:65026} }
\end{figure}

\begin{figure}
\epsscale{1.1}
\vspace*{0.5cm}
\plotone{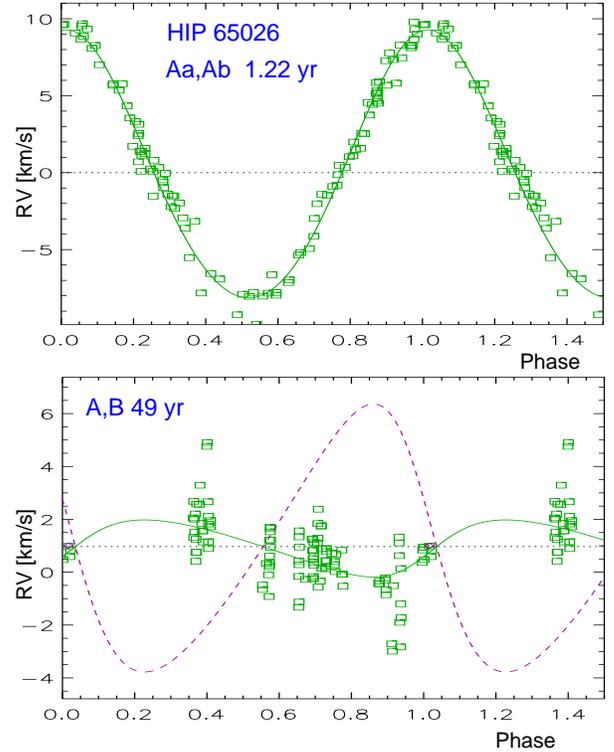}
\caption{RV curves  of the  inner (top) and  outer (bottom)  orbits of
  HIP~65026  from   the  combined   solution.   In  each   curve,  the
  contribution of other orbit is subtracted. The dotted line in the
  lower plot shows the expected RV curve of the component B. 
\label{fig:65026RV} }
\end{figure}

\begin{figure}
\epsscale{1.1}
\plotone{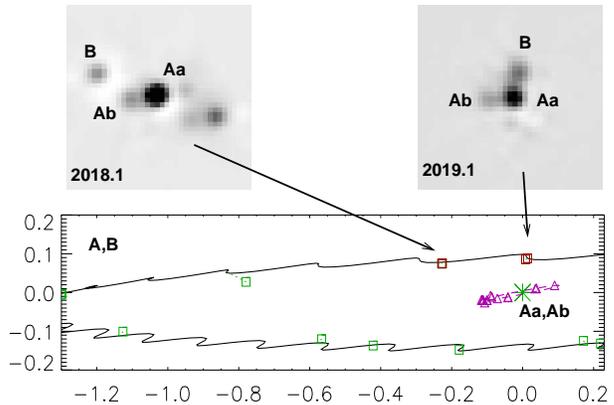}
\caption{Fragment of the  outer and inner orbits of  HIP~65026 and the
  reconstructed images of the triple  star recorded with the red NESSI
  camera  in  2018  (separations  0\farcs24 and  0\farcs10)  and  in 2019
  (separations   0\farcs09  and   0\farcs08),   when  the   previously
  unobserved part of the outer orbit was covered. The wavy line
  shows the wobble produced by the inner subsystem.
\label{fig:CHR193} }
\end{figure}

This triple system is known as HIP~65026, HD~115953, GJ~508, ADS~8861,
WDS J13198+4747,  and BD+48~2108.  Simbad gives the  spectral type M2V
and  contains  102  references.    The  {\it  Gaia}  DR2  parallax  is
109.98$\pm$0.83  mas  (distance  9\,pc),  and  the  proper  motion  is
$(+226.5, -23.8)$ \masyr.

The visual binary HU~644 A,B was discovered in 1904.  It has made
2.3 revolutions on  its 49 yr orbit, which is  now very well defined
(grade 2).  The inner subsystem CHR~193 Aa,Ab was discovered in 1992.3
at  0\farcs11  separation by  \citet{Hrt1994}  and  has been  measured
several  times since.  Independently,  the RV  variation was  found by
\citet{TS02}   (observations  in   1995,  1996,   and  2000)   and  by
\citet{Sperauskas2019} (2011 to 2018).   The latter authors computed a
spectroscopic orbit with  a period of 447 days  (1.22 yr) which refers
to the inner pair Aa,Ab. \citet{Beuzit2004} resolved the triple system
using adaptive optics and  announced a preliminary spectroscopic orbit
of  the inner  pair with  $P=450$  d that  has not  been published  so
far. Beuzit et al. wrote that the spectra are triple-lined.

As a first  step, the outer and inner  orbits were analyzed separately
(Figure~\ref{fig:65026}).  We  fit   the  outer  orbit using  all
available data, but then fix the outer period to 49.077 yr and fit the
remaining elements  using only speckle and adaptive  optics data. Both
orbits  have  a large  inclination.   The  inner  orbit uses  the  RVs
mentioned above  and additional 87  unpublished RVs measured  with the
CfA spectrometers  between 1995.5  and 2006.4.  Owing  to the  long RV
coverage, the inner period is very accurate.

The two orbits were then fitted simultaneously using the IDL code {\tt
  orbit3.pro}  \citep{TL2017}.  Only  interferometric  measurements of
the outer pair  are used, and its period,  determined by the long-term
visual coverage, is kept fixed.   The resulting elements are listed in
Table~\ref{tab:comborb}.   The   global  weighted  rms   residuals  in
position are 5 mas for  the inner and outer subsystems.  However, some
observations of the outer pair left unrealistically large residuals of
almost 30\,mas and  caused divergence of the combined  fit.  They were
carefully  examined  and re-weighted.   The  measurements in  1989.13,
2000.32, 2005.235, 2005.32, and 2013.29 were ignored, while some other
data were  assigned errors  larger than usual.   The RV  residuals are
0.72 \kms.   The results of the  joint solution are  very similar 
  (mostly within the errors)  to the orbits computed separately.  The
RV curves are presented in Figure~\ref{fig:65026RV}.  The combined fit
defines additional parameter, the wobble factor $f_{\rm Aa,Ab} = 0.425
\pm 0.025$  (ratio of  the wobble amplitude  to the inner  axis).  The
wobble amplitude is 49\,mas.  Figure~\ref{fig:CHR193} shows a fragment
of the  outer orbit with wobble and  the reconstructed high-resolution
images obtained at WIYN.  They cover previously unobserved part of the
outer orbit.

The  nodes of  both orbits  are  known from  the RVs,  allowing us  to
compute  the  angle  between  the orbital  angular  momenta  (relative
inclination) of 11\fdg3$\pm$1\fdg0.   The orbits are almost, but not
quite,  coplanar, and  have small  eccentricities.   Intriguingly, the
period ratio is 39.97$\pm$0.005, suggesting  that the orbits may be in
a weak  mean motion  resonance.  Such architecture  is found  in some
other low-mass  triple system  and hints at  their origin in  a common
disc \citep{twins}.

The {\it Gaia} parallax of 109.98$\pm$0.83 mas is likely biased by the
astrometric subsystem  with a period of  1.2 yr, not  accounted for in
the DR2 astrometric solution (the  large parallax error shows its poor
quality). The components' masses derived  from the orbits and the {\it
  Gaia} parallax do not quite  match the absolute magnitudes. We adopt
here the  parallax of 101\,mas computed  from the outer  orbit and the
estimated  masses  of  Aa, Ab,  and  B,  0.58,  0.42, and  0.48  \msun
~respectively (sum  1.51 \msun).   With this parallax,  the components
are located  on the main sequence.   To match this model,  the axis of
the  inner orbit  should be  larger than  measured by  4\%,  while its
formal error  is 1.8\%.  Considering  the difficulty of  measuring the
inner pair in this triple system, such disagreement is not alarming.

We determine  $V =  9.33, 10.83, 10.21$  mag for  Aa, Ab, and  B using
published and  new differential photometry together  with the combined
magnitude $V_{\rm tot}=8.76$ mag.  The wobble factor computed from the
above  masses  and magnitudes  is  $f_{\rm  Aa,Ab}=0.44$ for  resolved
measurements of Aa,B and $f_{\rm Aa,Ab}=0.24$ for photo-center measurements of
A,B in the  $V$ band; the measured wobble  factor is 0.42.  Similarly,
the wobble factor of the outer orbit would be 0.33 (resolved) and 0.07
(photo-center). \citet{Heintz1969}  determined the photo-center wobble
of  the outer  orbit  as $f_{\rm  A,B}  = 0.088$.  At  that time,  the
existence of  the one-year  subsystem Aa,Ab was  not known,  so Heintz
derived a biased parallax of 119$\pm$3.5 mas.

The masses,  period, and inclination of  the outer system  allow us to
compute the RV  amplitudes of 3.6 and 5.1  \kms for A (center-of-mass)
and B,  respectively. Yet,  the free fit  yields a much  smaller outer
amplitude $K_1 = 1.07$  \kms. We tentatively attribute the discrepancy
to blending with other components that reduces the amplitude.

The  study of  this  remarkable low-mass  triple  system will  further
benefit from  spectroscopy with a higher resolution  and, possibly, at
longer wavelengths to resolve spectrally all three stars and determine
their unbiased  RV amplitudes. High inclination of  the orbits favors
measurement  of the  orbital  parallax. Meanwhile,  future {\it  Gaia}
releases should  include orbital motion  in the astrometric  model,
delivering accurate astrometric orbits and unbiased parallax.

\subsection{The quadruple system HIP 85209}
\label{sec:85209}

\begin{figure}
\epsscale{1.1}
\plotone{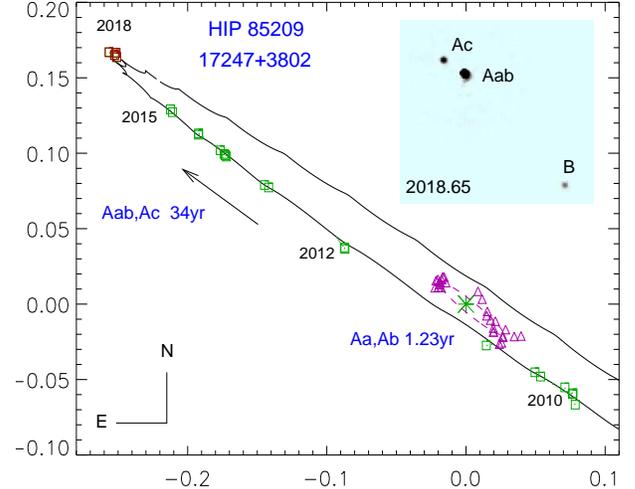}
\caption{Visual orbits of HIP 85209 (WDS J17247+3802). Fragment of the
  outer orbit  and the  inner orbit are  plotted to scale.  The insert
  shows the reconstructed  speckle image in 2018.65 in  the red channel
  with components Ac and B (the inner pair Aa,Ab is unresolved).
\label{fig:85209} }
\end{figure}

\begin{figure}
\epsscale{1.1}
\plotone{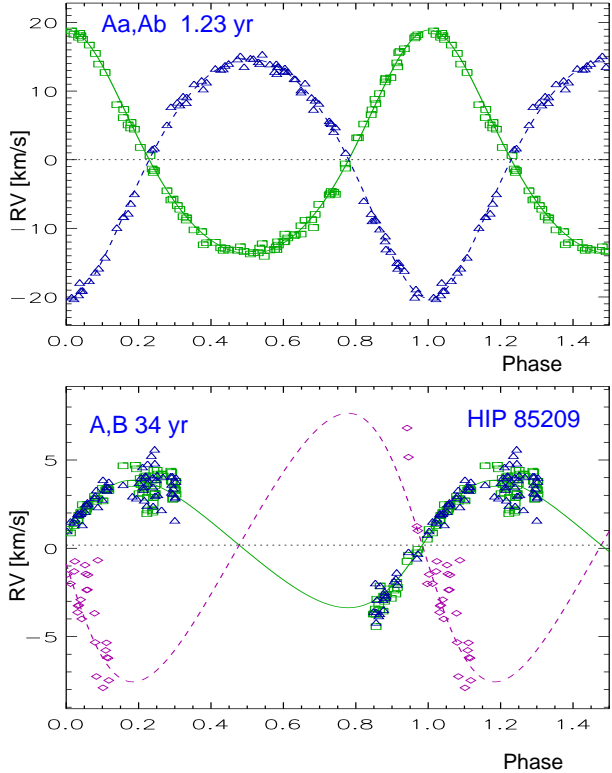}
\caption{RV curves  of the  inner (top) and  outer (bottom)  orbits of
  HIP~85209  from   the  combined   solution.   In  each   curve,  the
  contribution of other orbit is subtracted. 
\label{fig:85209RV} }
\end{figure}

17247+3802  (HIP  85209, HD  157948)  is  a  quadruple system  of  3+1
hierarchy  located at 50  pc from  the Sun.   The outer  1\farcs8 pair
COU~1142 A,B  is known since  1974; its estimated period  is $\sim$500
yr. On  the other  hand, this is  a double-lined  spectroscopic binary
with a period of  448.6 days \citep{Goldberg2002}.  This pair, denoted
in  WDS as  HSL~1 Aa,Ab,  has been  resolved by  both  the fine-guider
sensors of  the Hubble Space Telescope \citep{Hsl2002,Horch2006}  and by speckle
interferometry,  allowing  \citet{Hor2015}  to  determine  its  visual
elements.   Speckle  interferometry   also  revealed  an  intermediate
subsystem  Aab,Ac (HSL~1  Aab,Ac)  at 0\farcs3  separation. Its  first
orbit with $P=122$ yr and a large eccentricity of $e=0.72$ computed by
\citet{Rbr2018} does  not agree with the latest  observations that
cover now 17 years and suggest a quasi-circular orbit with $P \approx 38$ yr.

To gain a better understanding of this system, we fitted the inner and
middle  orbits   simultaneously  using  the   latest  measurements  by
\citet{Horch2019} in addition to  the published and our own astrometry
and  the RVs from  \citet{Goldberg2002}. 
Their  measurements, covering
the period from 1982.5 to 1987.5, are reprocessed and extended to 2009.
Furthermore, the spectra from TRES obtained from 2009.4 to 2015.4 were
processed by {\tt  TRICOR}, extracting the RVs of Aa,  Ab, and Ac. The
RV data therefore  cover almost entire outer period  of 34 yr,
constraining the outer orbit.  The elements of both orbits
fitted jointly are given in Table~\ref{tab:comborb} and illustrated in
Figures~\ref{fig:85209}  and   \ref{fig:85209RV}.   The inner spectroscopic
elements differ  only slightly from the published  orbit; the weighted
rms residuals in RV are 0.19, 0.37,  and 1.92 \kms for Aa, Ab, and Ac,
respectively.  The  resulting masses  of Aa and  Ab are 0.91  and 0.85
\msun.   The  {\it  Gaia}  parallax of  20.676$\pm$0.11\,mas  for  the
component B (it is more accurate than the parallax of A) and the inner
orbit lead to the mass sum of 1.84 \msun ~that agrees with the
spectroscopic masses. The outer
orbit correspond to  the mass sum of 2.77 \msun. However,
the RV amplitudes in the outer orbit imply smaller masses of
1.20 and 0.57 \msun for Aa+Ab and Ac, respectively. This means that the
 RV amplitudes in the outer orbit are slightly under-estimated,
possibly because the RVs measured by {\tt TRICOR} are biased by
blending in the triple-lined spectra. 

The   components  Aa   and   Ab   are  similar   in   both  mass   and
brightness. Using  the {\it Gaia}  photometry $V_{\rm A} =  8.08$ mag,
$\Delta  V   _{\rm  Aa,Ab}  =  0.4$  mag   \citep[in  rough  agreement
  with][]{Horch2019}, and $\Delta m  _{\rm Aab,Ac} = 3.0$ mag measured
here, we  get the individual $V$  magnitudes of 8.72,  9.12, and 11.17
mag  for Aa,  Ab, and  Ac. Their  absolute magnitudes  and  the masses
measured here match standard  relations for main-sequence stars.  This
star  has  been  considered  to  be  metal-poor.  However,  given  the
composite nature  of the spectrum,  its standard analysis may  lead to
biased results. The kinematics does not distinguish HIP~85209 from the
population of Galactic disk.

The inner  and middle orbits  are seen nearly edge-on;  their relative
inclination is 12\fdg0$\pm$3\fdg0. The  position angle of the pair A,B
 (221\degr)  is close to  the position angle  of both inner orbits, while the
observed motion of A,B is mostly radial  (increasing separation), suggesting
that the orbit of A,B can be oriented approximately in the same
plane. Therefore, this is a 3+1 ``planetary'' hierarchy with
quasi-coplanar and quasi-circular orbits.

\subsection{New binaries}

\begin{deluxetable}{l l cc  cc }    
\tabletypesize{\scriptsize}     
\tablecaption{Newly resolved pairs
\label{tab:new}          }
\tablewidth{0pt}                                   
\tablehead{                                                                      
\colhead{WDS} & 
\colhead{HR} & 
\colhead{$\rho$} & 
\colhead{$\theta$} &  
\colhead{$\Delta m$(562)} &
\colhead{$\Delta m$(716)} 
\\
 & & 
\colhead{(arcsec)} & 
\colhead{(deg)} & 
\colhead{(mag)} &
\colhead{(mag)} 
}
\startdata
00585+6621 &  273 & 1.47 & 58  & 6.1 & 5.7 \\ 
01029+4121 &  290 & 0.10 & 143 & 2.0 & 1.5 \\
04551+5516 & 1555 & 0.05 & 155 & 0.2 & 0.2 \\ 
13057+3548 & 4943 & 0.22 & 63  & 4.1 & 4.2 
\enddata 
\end{deluxetable}


Four   reference  stars  were   unexpectedly  resolved   as  binaries.
Table~\ref{tab:new} lists these  objects, while all their measurements
are found  in the main  Table~\ref{tab:measures}.  Three stars  with a
large  $\Delta  m$  still   provide  useful  references  in  the  data
processing because the binary  signature is automatically removed from
the reference power spectrum by our pipeline.

Two newly resolved pairs are in fact triple systems. The new 1\farcs47
companion to  HR~273 (HIP 4572,  HD~5550, A0III, 110\,pc) is  found in
the {\it  Gaia} DR2 at  a similar position, 1\farcs455  and 58\fdg4.
Considering the  PM of  45 \masyr, we  conclude that this  companion is
physical (co-moving).  Meanwhile, the  star A is a chemically peculiar
double-lined spectroscopic binary with a period of 6.82 d according to
\citet{Carrier2002}.  Therefore,  this is a  new triple system  hosting a
close inner pair.

HR~1555  (HIP~22854, HD~30958,  ADS~3508, B9.5V,  236\,pc) is  a known
visual  binary BU~1187  with  a separation  of  12\farcs7. {\it  Gaia}
astrometry  confirms  that this  binary  is  physical  (common PM  and
parallax). Its  main component  A is resolved  here into a  tight pair
Aa,Ab; the  separation of  47\,mas implies a  short orbital  period of
$\sim$15\,yr. Indirectly,  this pair  is confirmed by  the astrometric
acceleration, $\Delta \mu  = (-2.6, -3.6)$ \masyr. The  orbit of Aa,Ab
can be determined within several years.

\subsection{Spurious pairs}
\label{sec:bogus}

\begin{deluxetable}{l l c c  c}    
\tabletypesize{\scriptsize}     
\tablecaption{Unconfirmed subsystems
\label{tab:bogus}          }
\tablewidth{0pt}                                   
\tablehead{                                                                      
\colhead{WDS} & 
\colhead{Discoverer} & 
\colhead{$\rho$} & 
\colhead{Last} &
\colhead{$P^*$}
\\
 & 
\colhead{code} 
& 
\colhead{(arcsec)} & 
\colhead{(yr)} &  
\colhead{(yr)}   
}
\startdata
02132+4414 & CHR 5  & 0.2  & 1987 & 80 \\ 
03127+7133 & SCA 171 Aa,Ab & 0.4  & 2009 & 60 \\ %
03492+2408 & CHR 125 Aa,Ab & 0.2    & 1991 & 60 \\ %
04184+2135 & LMP 52 Aa,Ad &  0.3  & 2000 & 40 \\ 
04493+3235 & CHR 19  &  0.04  & 1984 & 5 \\ 
09068+4707 & COU 2687 &  0.45  & 1993 & 100 \\ 
09188+3648 & CHR 173 Ba,Bb & 0.24 & 2004 & 15 \\ 
10454+3831 & CHR 191 A,B & 0.36  & 1991 & 15 \\ 
16238+6142 &  CHR 138 Aa,Ab & 0.2  & 1990 & 90 \\ 
17491+5047 & CHR 65 Aa,Ab & 0.12 & 1985 & 11 \\ 
18443+3940 & CHR 77 Ca,Cb & 0.10 & 2005 & 20 \\ 
19111+3847 & STF 2481 Aa,Ab &  0.1  & 2012 & 10 \\ 
22139+3943 & BNU 8 Aa,Ab & 0.19  & 2005 & 60  \\ 
23439+3232 & BAG 30 Ba,Bb & 0.2 & 2000 & 10 \\ 
23460+4625 & MCA 75 Aa,Ab & 0.2  & 2003 & 75 
\enddata 
\end{deluxetable}

Some subsystems  are not confirmed  by subsequent observations.  It is
important to recognize  these cases for cleaning up  the statistics of
hierarchical multiplicity.  However, it is  much easier to  discover a
subsystem  than   to  prove  its  non-existence   because  a  negative
observation  does   not  necessarily   mean  that  the   subsystem  is
spurious. It could have closed  down below the resolution limit or the
contrast  may exceed  the  dynamic  range of  the  technique. A  crude
estimate of an orbital period  $P^*$ based on the projected separation
(assuming it equals the semimajor axis) helps to evaluate the veracity
of subsystems.  We also  considered other evidence such as astrometric
acceleration, RV variability, and observations of outer subsystems.

Table~\ref{tab:bogus} lists  likely spurious subsystems  unresolved at
WIYN.   Its first  two columns  provide the  WDS codes  and discoverer
designations. The  latest measured separations  and the epochs  of the
last reported resolution  are given in the following  two columns. The
last column gives the period estimate. 

The existence of inner and outer subsystems with comparable
separations is highly improbable because such triples would be
dynamically unstable, unless their configuration is a result of
projection. When the outer orbit is known, the projection is no longer
relevant and we can confidently rule out spurious inner pairs. This is
the case of 03127+7133 (outer period 583 yr according to our updated
orbit), 10454+3831 (outer period 326 yr), and 22139+3943 (outer period
125 yr). Another spurious ``trapezium'' triple is 23460+4625.

\section{Summary}
\label{sec:sum}

\begin{figure}
\epsscale{1.1}
\plotone{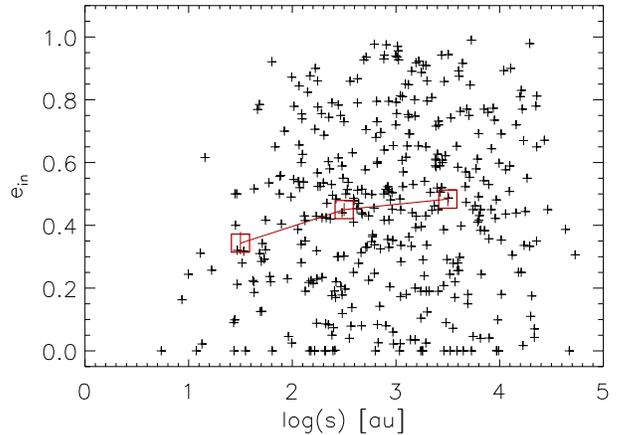}
\caption{Eccentricity of inner visual orbits in hierarchical systems
  vs. projected separation $s$ of their outer components. The red squares with
  error bars plot mean eccentricities and their errors in the separation
  bins of one dex.
\label{fig:inner} }
\end{figure}

Our  observations make  an  incremental contribution  to  the data  on
hierarchical multiple  systems. Astrometric measurements  at WIYN will
help  to determine  orbits of  many  subsystems in  the future,  while
several orbits  are computed here.  Mostly they refer to  low-mass and
nearby hierarchies. Our  work also contributes differential photometry
in two passbands. We discovered serendipitously two new triple systems
and confirmed the spurious nature of some previous resolutions.

Low-mass hierarchies are  often found in ``planetary'' configurations,
where  the inner  and outer  orbits are  approximately  co-planar, the
pairs are  co-rotating, and  the orbital eccentricities  are moderate,
like in  HD 91962 \citep{planetary}.  Two such  systems, HIP~65026 and
85209,  are studied  here in  detail.  We  accurately  measured mutual
orbit  inclinations   in  both  systems,   proving  their  approximate
co-planarity.   They  also     have  moderate   eccentricities,  
presumably being sculpted by dissipative evolution in accretion disks.
Despite   similarities  with   planetary  systems,   however,  stellar
hierarchies have less coplanar and more eccentric orbits.

However, not all subsystems  have quasi-circular orbits.  Quite to the
contrary, nearly  half of the orbits in  Table~\ref{tab:vborb}  (13 out
of 23)  have eccentricities exceeding 0.5. Large eccentricities suggest
an   important  role   of  dynamical   interactions  in   shaping  the
architecture of  these hierarchies.  In this  case, the eccentricities
are  expected to  follow the  thermal distribution  $f(e)=2e$  and the
mutual  orbit   inclinations  in  triple  systems   should  be  random
(uncorrelated). Nearly  orthogonal orbits should  further increase the
eccentricities of  inner subsystems through Kozai-Lidov oscillations. 

It  is likely  that the  role of  dynamical interactions  increases at
larger   spatial  scales.    To   probe  this   idea,   we  plot   in
Figure~\ref{fig:inner}  eccentricities  of   inner  visual  orbits  in
hierarchical  systems  versus   projected  separation $s$  of  the  outer
components,  based on  the  current version  of  the MSC.  We made  no
attempt  to filter  the  quality of  397  visual orbits  in this  plot
(including the orbits determined here) and caution that the MSC 
is burdened  by numerous selection effects.  Nevertheless,  a trend of
smaller inner eccentricities at  projected outer separations $s < 100$
au is  emerging.   As  noted by \citet{Tok2017},  compact hierarchies
also tend to have more aligned orbits. Further accumulation of data on
orbits in  hierarchical stellar  systems, especially in  compact ones,
will help to clarify these trends.


\acknowledgments 

Observations in  the paper  made use of  the NN-EXPLORE  Exoplanet and
Stellar Speckle Imager (NESSI). NESSI was funded by the NASA Exoplanet
Exploration Program and the NASA Ames Research Center. NESSI was built
at the  Ames Research  Center by Steve  B. Howell, Nic  Scott, Elliott
P. Horch, and Emmett Quigley.   This project is partially supported by
the  NASA award  \#1598596. B.~Mason  has kindly  commented  the draft
version of our paper.

This work  used the  SIMBAD service operated  by Centre  des Donn\'ees
Stellaires  (Strasbourg, France),  bibliographic  references from  the
Astrophysics Data  System maintained  by SAO/NASA, and  the Washington
Double Star  Catalog maintained  at USNO.  This  work has made  use of
data   from   the   European   Space   Agency   (ESA)   mission   Gaia
(\url{https://www.cosmos.esa.int/gaia})  processed  by  the  Gaia  Data
Processing      and     Analysis      Consortium      (DPAC,     {\url
  https://www.cosmos.esa.int/web/gaia/dpac/consortium}). Funding for the
DPAC  has been provided  by national  institutions, in  particular the
institutions participating in the Gaia Multilateral Agreement.

{\it Facilities:}  \facility{WIYN}.



\end{document}